\documentclass[preprint]{aastex}
\usepackage{amsmath}
\usepackage{amssymb}
\usepackage{seqsplit}
\usepackage{natbib}
\usepackage{epstopdf} 
\usepackage{graphicx}
\usepackage{wrapfig}
\usepackage{lscape}
\usepackage{rotating}

\begin{document}

\bibliographystyle{apj} 

\begin{titlepage}

\title{Quantitative Spectroscopy of Blue Supergiants in Metal-Poor Dwarf Galaxy NGC 3109}
\author{Matthew W. Hosek Jr.\altaffilmark{1}, Rolf-Peter Kudritzki\altaffilmark{1}, Fabio Bresolin\altaffilmark{1}, Miguel A. Urbaneja\altaffilmark{2},  Christopher J. Evans\altaffilmark{3}, Grzegorz Pietrzy\'{n}ski\altaffilmark{4,5},Wolfgang Gieren\altaffilmark{4}, Norbert Przybilla\altaffilmark{2}, and Giovanni Carraro\altaffilmark{6} }
\altaffiltext{1}{Institute for Astronomy, University of Hawaii, 2680 Woodlawn Drive, Honolulu, HI 96822, USA; mwhosek@ifa.hawaii.edu, kud@ifa.hawaii.edu, bresolin@ifa.hawaii.edu}
\altaffiltext{2}{Institute for Astro and Particle Physics, Innsbruck University, Austria; Miguel.Urbaneja-Perez@uibk.ac.at, Norbert.Przybilla@uibk.ac.at}
\altaffiltext{3}{UK Astronomy Technology Centre, Royal Observatory, Blackford Hill, Edinburgh, UK; chris.evans@stfc.ac.uk}
\altaffiltext{4}{Departamento de Astronom\'{i}a, Universidad de Concepci\'{o}n, Casilla 160-C, Concepci\'{o}n, Chile; pietrzyn@astrouw.edu.pl, wgieren@astro-udec.cl}
\altaffiltext{5}{Also at: Warsaw University Observatory, Al. Ujazdowski 4, 00-478 Warsaw, Poland}
\altaffiltext{6}{European Southern Observatory, La Silla Paranal Observatory, Chile; gcarraro@eso.org}

\begin{abstract}

We present a quantitative analysis of the low-resolution ($\sim$4.5~\AA) spectra of 12 late-B and early-A blue supergiants (BSGs) in the metal-poor dwarf galaxy NGC~3109. A modified method of analysis is presented which does not require use of the Balmer jump as an independent T$_{eff}$ indicator, as used in previous studies. We determine stellar effective temperatures, gravities, metallicities, reddening, and luminosities, and combine our sample with the early-B type BSGs analyzed by \citet{E07} to derive the distance to NGC 3109 using the Flux-weighted Gravity-Luminosity Relation (FGLR). Using primarily Fe-group elements, we find an average metallicity of [\={Z}] = -0.67 $\pm$ 0.13, and no evidence of a metallicity gradient in the galaxy. Our metallicities are higher than those found by \citet{E07} based on the oxygen abundances of early-B supergiants ([\={Z}] = $-$0.93 $\pm$ 0.07), suggesting a low $\alpha$/Fe ratio for the galaxy. We adjust the position of NGC 3109 on the BSG-determined galaxy mass-metallicity relation accordingly and compare it to metallicity studies of HII regions in star-forming galaxies. We derive an FGLR distance modulus of 25.55 $\pm$ 0.09 (1.27 Mpc) that compares well with Cepheid and tip of the red giant branch (TRGB) distances. The FGLR itself is consistent with those found in other galaxies, demonstrating the reliability of this method as a measure of extragalactic distances.

\end{abstract}
\maketitle
\end{titlepage}

\section{Introduction}

The extreme brightness of blue supergiants (BSGs), a short post-main sequence evolutionary stage of 12~M$_{\odot}$ to 40~M$_{\odot}$ stars, makes it possible to obtain resolved spectra of individual BSGs out to 10 Mpc with current instrumentation. As such, BSGs are ideal tools to obtain crucial information about the chemical composition of nearby galaxies and provide insight to their chemical evolution \citep{K08,K12}. Often galaxy metallicities are studied through the spectroscopy of HII regions, which has been widely applied to examine radial abundance gradients of spiral galaxies \citep{VC92, Z94, P04} and the galaxy mass-metallicity relation \citep{L79, T04, A13}. However, this approach is limited by its reliance on empirical ``strong line'' analysis methods, which have been shown to yield significantly different absolute metallicities depending on what calibration is used \citep{Ke08, B09}. Even in cases where metallicities can be measured more directly using the weak auroral lines, HII region studies might be affected by systematic uncertainties difficult to assess, such as oxygen depletion on to dust grains and a possible detection bias towards lower abundances at high metallicities \citep{ZB12}. BSGs thus provide a valuable independent measure of galaxy metallicity.

In addition, it has been shown that BSGs can be used as distance indicators through the Flux-weighted Gravity-Luminosity Relation (FGLR, Kudritzki et al. 2003, 2008). This relation correlates stellar gravity and effective temperature, which can be derived from the stellar spectrum, to the absolute bolometric magnitude. The FGLR is advantageous in that it is free of uncertainties caused by interstellar reddening, since the reddening is determined during the spectral analysis. In addition, a potential metallicity dependence of the FGLR, if present, can be accounted for since metallicity is also determined independently by the analysis of each object. This is especially valuable in light of recent efforts to establish the Hubble constant \emph{H$_{0}$} to an accuracy better than 5\%, which would greatly constrain cosmological parameters without having to invoke assumptions about the geometry of the universe \citep{KU12, Riess11}. FGLR distances found for WLM \citep{U08}, M33 \citep{U09}, and M81 \citep{K12} have been found to be consistent with distances determined by other methods, demonstrating the reliability of the method. 

We conduct a spectroscopic study of 12 late-B and early-A type BSGs in NGC~3109, a Magellanic SBm galaxy at the edge of the Local Group \citep{dV91,Bergh99}. With M$_{V}$ = -14.9 \citep{Mcc12}, the galaxy is the most luminous member of the NGC~3109 group, which according to \citet{Tully06} is the ``nearest distinct structure of multiple galaxies to the Local Group''. Recent work by \citet{ST13} and \citet{B13} indicate that the members of this group form a $\sim$1070 kpc filamentary structure created by tidal interaction or filamentary accretion. Our purpose is two-fold: to determine the FGLR and calculate a corresponding distance to NGC~3109, and to evaluate its average metallicity using multiple Fe-group element species. An analysis of eight early B-type BSGs by Evans et al. (2007, hereafter E07) found the oxygen abundance of NGC~3109 to be approximately 1/10 of solar, a result consistent with HII regions studies using auroral lines \citep{Le03, Le03b, P07}. If these values reflect the overall metallicity, NGC~3109 will be the lowest metallicity object for which an FGLR has been constructed, allowing us to investigate the metallicity dependence of the relation and compare with stellar evolution theory. Since we fit metal lines of multiple elements, in particular Fe-group elements, our metallicities will more closely resemble the overall stellar metallicities and are not restricted to oxygen as a proxy of stellar metallicity as in the case of the HII region studies and the work of E07. On the other hand, a comparison with the oxygen abundances obtained in these studies will provide insight to the $\alpha$/Fe abundance ratio, a key parameter in constraining the chemical evolution and star formation history of the galaxy. 

\section{Method}

\subsection{Observations and Sample}
We analyze low-resolution (R $\sim$1000) spectra of late-B and early-A supergiants obtained by E07 in order to derive the effective temperature (T$_{eff}$), surface gravity (log \emph{g}), and average metallicity ([Z] = log Z/Z$_{\odot}$) of each object. A large sample (91 objects) of possible BSGs was identified by E07 using the V- and I-band photometry of \citet{Pi06}. These stars were observed with the FORS2 spectrograph at the VLT on 2004 February 24 and 25, operated in the moveable slits (MOS) mode using the 600B grism in the blue and the 1200R grism in the red. Our study uses the flux-normalized blue-region spectra (3650-5500~\AA), which exhibit a FWHM resolution of $\sim$4.5~\AA.  The observed widths of the metal lines are established by the resolution of the spectrograph rather than intrinsic properties of the star. For details regarding the extraction and reduction of these spectra, as well as the spectral classification of the individual objects, we refer the reader to E07. 

\begin{figure}
\begin{center}
\includegraphics[scale=0.3]{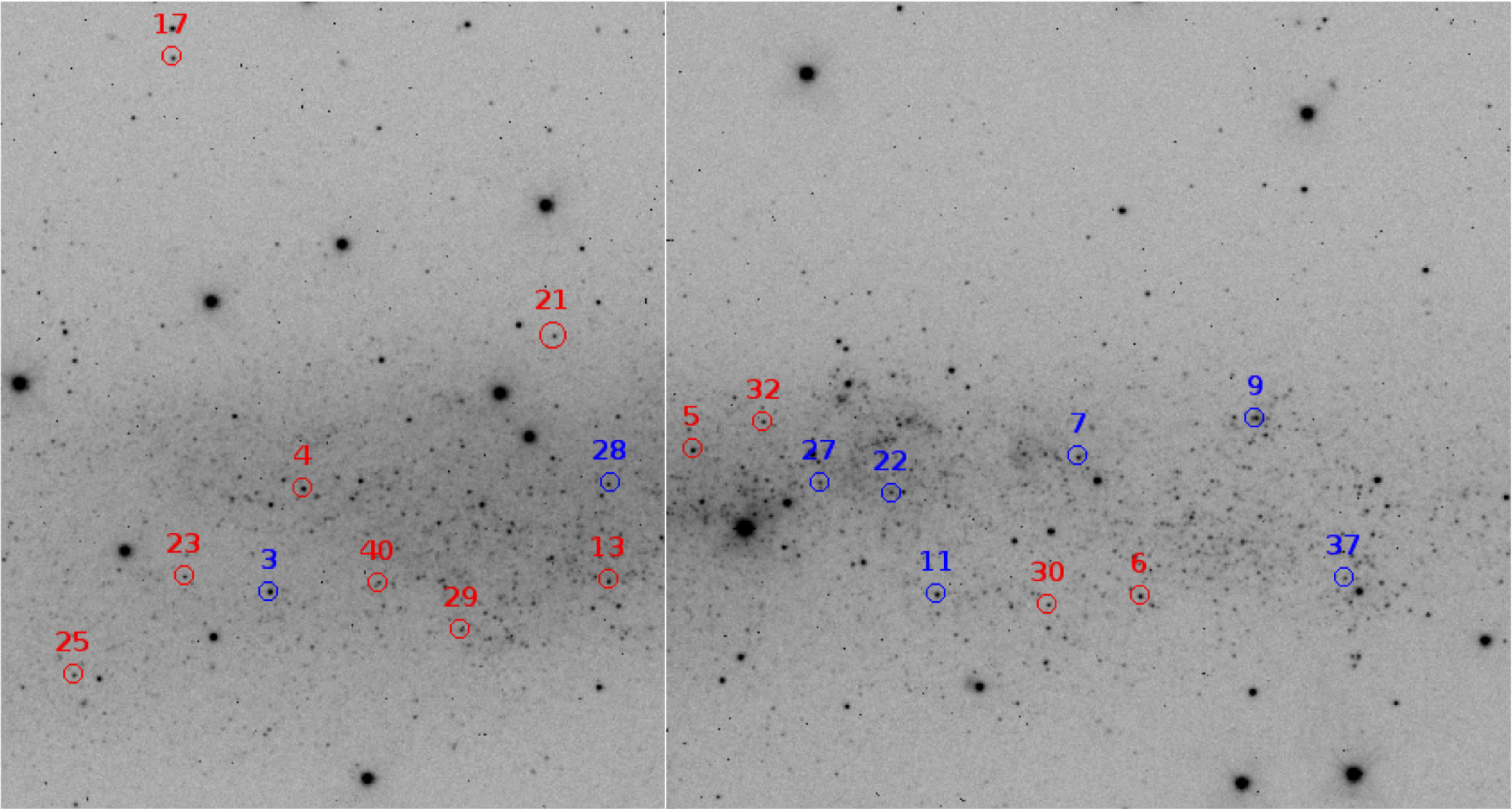}
\caption{The positions of the BSGs analyzed by this study (red) and Evans et al. 2007 (blue). This V-filter image of NGC~3109 was taken using the Warsaw 1.3m telescope at Las Campanas Observatory in 2003 May.}
\label{NGC3109_composite}
\end{center}
\end{figure}

Our sample of late-B and early-A supergiants consists of 12 of the E07 objects (Table \ref{Myobjects}, Fig. \ref{NGC3109_composite}). An average signal-to-noise ratio (S/N) is calculated for each spectrum from multiple line-free regions across the full wavelength range. While more BSGs are present in the E07 sample, our analysis method only proved successful in constraining stellar parameters for stars with high S/N spectra ($\ge$ $\sim$50). 

\subsection{Spectral Analysis}

It has been shown that the effective temperature of late-B and early-A type stars can be very accurately determined using the ionization equilibria of weak metal lines such as O I/II, Mg I/II, and N I/II \citep{Pr06,F12}. However, these lines cannot be reliably measured in low resolution spectra and thus an alternative method must be used. While the metal lines are sensitive to temperature, a degeneracy between temperature and metallicity makes it difficult to constrain T$_{eff}$ in this way (see Fig. 2 of Kudritzki et al. 2008). Previous studies of low-resolution BSG spectra have broken the degeneracy via the Balmer jump ($\sim$3646~\AA), which is also sensitive to temperature but largely independent of metallicity \citep{K08,K12, U08}. Unfortunately, a poor flux calibration in this region for the NGC~3109 spectra prevents a similar approach in this study, and is the reason why the late-B and early-A type stars were not analyzed by E07. Instead, we present a new spectral synthesis method that takes advantage of the fact that different metal lines react differently to changes in temperature and metallicity, partially breaking the temperature-metallicity degeneracy. For hotter stars, additional information provided by the temperature-dependent He I lines allow us to fully break the degeneracy. We discuss the details below. 

\subsubsection{The Model Atmosphere Grid}
The model atmosphere grid used in this study contains LTE line-blanketed atmospheres with the detailed non-LTE line formation calculations of Przybilla et al. (2006), as discussed in Kudritzki et al. (2008). The grid contains temperatures from 7900-15,000~K (spaced at increments of 250~K from 7900-10,000~K and 500~K from 10,000-15,000~K) and gravities between log \emph{g} = 0.8 and 3.0 (cgs, spaced at increments of 0.05 dex), where the lowest log \emph{g} value at each T$_{eff}$ is established by the Eddington limit. Metallicities are calculated relative to solar abundance for the following values of [Z] = log (Z/Z$_{\odot}$): $-$1.30, $-$1.15, $-$1.00, $-$0.85, $-$0.70, $-$0.60, $-$0.50, $-$0.30, $-$0.15, 0.00, 0.15, 0.30, and 0.50 dex. For each grid point, a microturbulence velocity $v_{t}$ is adopted based on the observed relationship between $v_{t}$ and log \emph{g} found in high-resolution studies of A-type BSGs in the Milky Way \citep{Pr02, Pr06, F12, V95a, V95b, V00, V01, V03, K04}. Figure 1 of \citet{K08} shows the distribution of the model atmosphere points in the (T$_{eff}$, log \emph{g}) plane together with the information about the microturbulence velocities used. 

Our original model grid assumes that He abundance increases slowly with metallicity (see Table \ref{Normal_He}) as indicated by HII regions and nucleosynthesis studies \citep[e.g.][]{PP74, Pa92}. While these values reflect the average He abundances of the young stellar populations examined in these studies well, detailed high-resolution studies of late B- and early A-type BSGs in the SMC \citep{Sch10} and the Milky Way (Firnstein and Przybilla 2012) reveal that many objects have He abundances higher than those predicted for their metallicities. This He enhancement is usually accompanied by a strong increase of nitrogen and a depletion of carbon that is interpreted as the result of rotationally-induced mixing during the advanced stages of stellar evolution. The grid takes this additional enrichment effect into account. We adopt a He abundance of $y~=~\frac{n_{He}}{n_{H} + n_{He}}$~=~0.12 at solar metallicity, as is found for main sequence B stars in the solar neighborhood \citep{NP12}. This value is supported by \citet{F12}, who find He abundances between y~=~0.11 and 0.13 for nearly all of the 35 A- and B-type supergiants they studied.

However, the situation is somewhat more ambiguous at the lower metallicity of the SMC. Of the 31 objects studied by \citet{Sch10}, 18 have a helium abundance between y = 0.08 and 0.10, which is why we adopt y = 0.09 close to SMC metallicity. However, the remaining 13 objects have higher helium abundances, mostly between y = 0.11 and 0.13. Because the He lines are an important part of our analysis (at least for the hotter objects), the assumptions about the He abundance could affect our results. To test this, we create a second model atmosphere grid identical to the first but with a constant enhanced He abundance y~=~0.13 for each metallicity and move forward in our analysis using both grids independently. In this way, we can assess the influence of the adopted helium abundance on the determination of effective temperature and metallicity in the anticipated low metallicity regime of NGC~3109. As shown below, the effects turn out to be small.

\subsubsection{Gravities, Temperatures, and Metallicities}

\begin{figure}
\begin{center}
\includegraphics[scale=0.4]{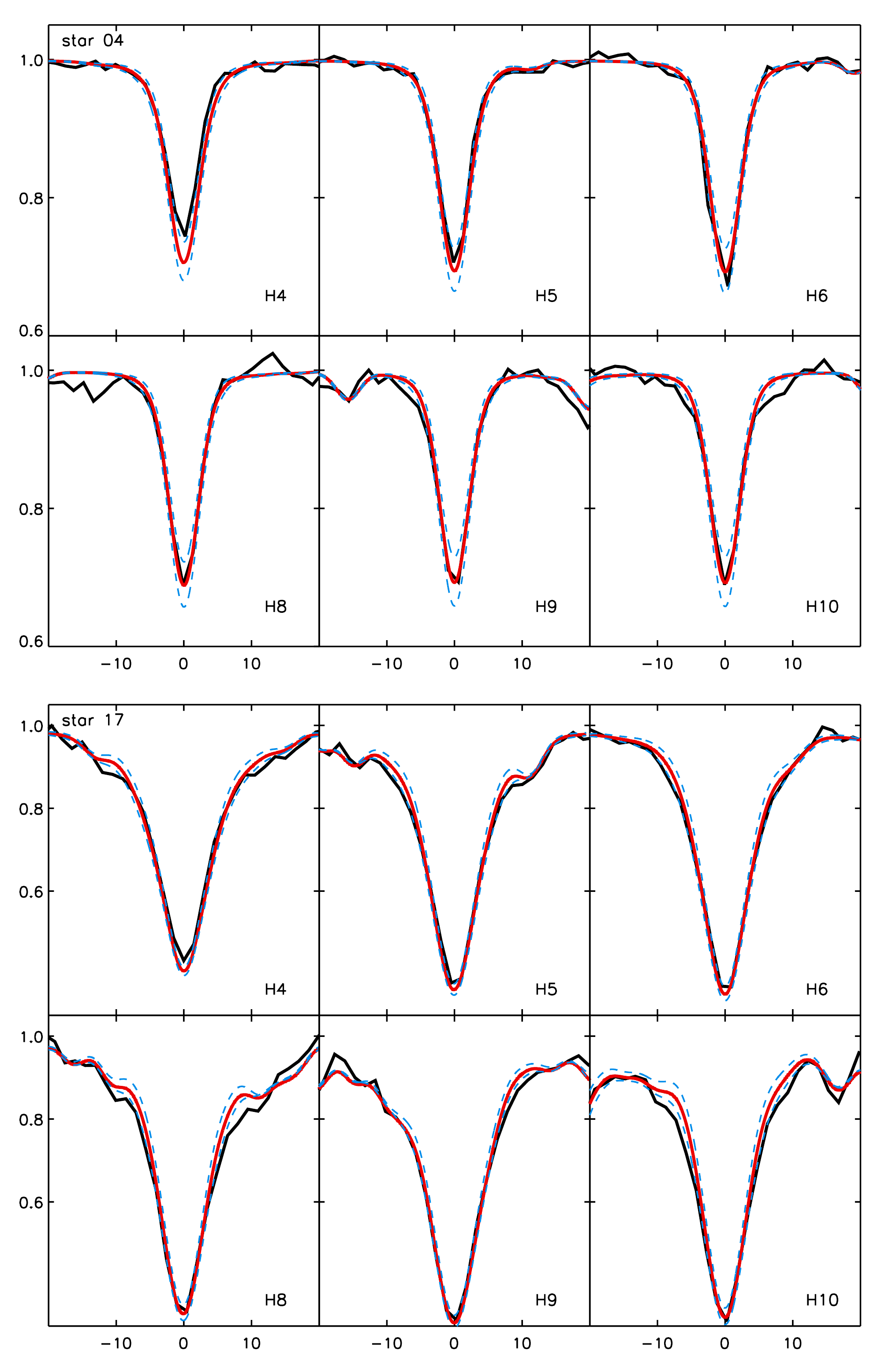}
\caption{Balmer line fits for star 4 (top 2 rows) and star 17 (bottom 2 rows), adopting a normal He abundance. The fits for the enhanced He abundance models are of similar quality. The black line is the observed spectrum, red line the best-fit model, and the blue dashed lines are the best-fit models with log g increased and decreased by 0.1 dex. The Balmer line fits for the other stars can be found in the appendix.}
\label{Balmer}
\end{center}
\end{figure}

Following Kudritzki et al. (2008, 2012), the first step in our analysis is to use the Balmer lines to constrain the possible values of log \emph{g} for each object. The low-resolution profiles and equivalent widths of the Balmer lines depend primarily on effective temperature and gravity, and can be fitted equally well by models with low temperature and low gravity as well as high temperature and high gravity. This establishes a Balmer line fit isocontour in the (T$_{eff}$, log \emph{g}) plane along which the observed profile and equivalent width agrees with the models (see Fig. 5 and 8 of Kudritzki et al. 2008, or Fig. 4 of Kudritzki et al. 2012). We evaluate multiple Balmer lines (H5, H6, H8, H9, and H10) to determine the gravity and assess accuracy. H4 is not used since it is usually contaminated by emission from stellar winds and surrounding HII regions, which can also affect H5 in extreme cases. H7 is frequently blended with interstellar Ca II absorption and therefore is not used in the analysis, as well. In cases where the H4 or even H5 fits indicate lower gravities than the remaining higher Balmer lines, we attribute this to the effects of winds or HII emission and rely instead on the higher Balmer lines. The log \emph{g} fit is typically good to $\sim$0.05-0.1 dex at fixed T$_{eff}$ (Fig. \ref{Balmer} and Appendix). Since T$_{eff}$ is not yet determined, the final value of the gravity is still unknown. However, the resulting Balmer line fit isocontour in the (T$_{eff}$, log \emph{g}) plane restricts the log \emph{g} parameter space, greatly reducing the number of models to evaluate when determining T$_{eff}$ and [Z].

With the gravities constrained, we use a $\chi^{2}$ analysis with the model atmosphere grid to simultaneously fit T$_{eff}$ and [Z] for our objects. We define a set of spectral windows for the late-B and early-A BSGs which are free of Balmer lines, nebular/interstellar contamination, and can be easily matched to the synthetic spectra (Table \ref{Spectral_win}). These windows contain lines of various metal species (Fe I/II, Ti II, Cr II, Mg II, etc) and He I (typically only present for the hotter late-B type objects). Each spectral window is inspected and boundaries adjusted to avoid cosmic rays and other spectral defects which could affect the analysis. To accurately determine the continuum level of the windows, we define a region near each boundary that is free of spectral lines and take the median value as the continuum for that edge. A discrepancy between the boundary continuum levels is likely caused by a residual spectral slope after the flux-normalization, and is fit to first order by linear interpolation. We find these discrepancies to be small, if present at all. 

\begin{figure}
\begin{center}
\includegraphics[scale=0.45]{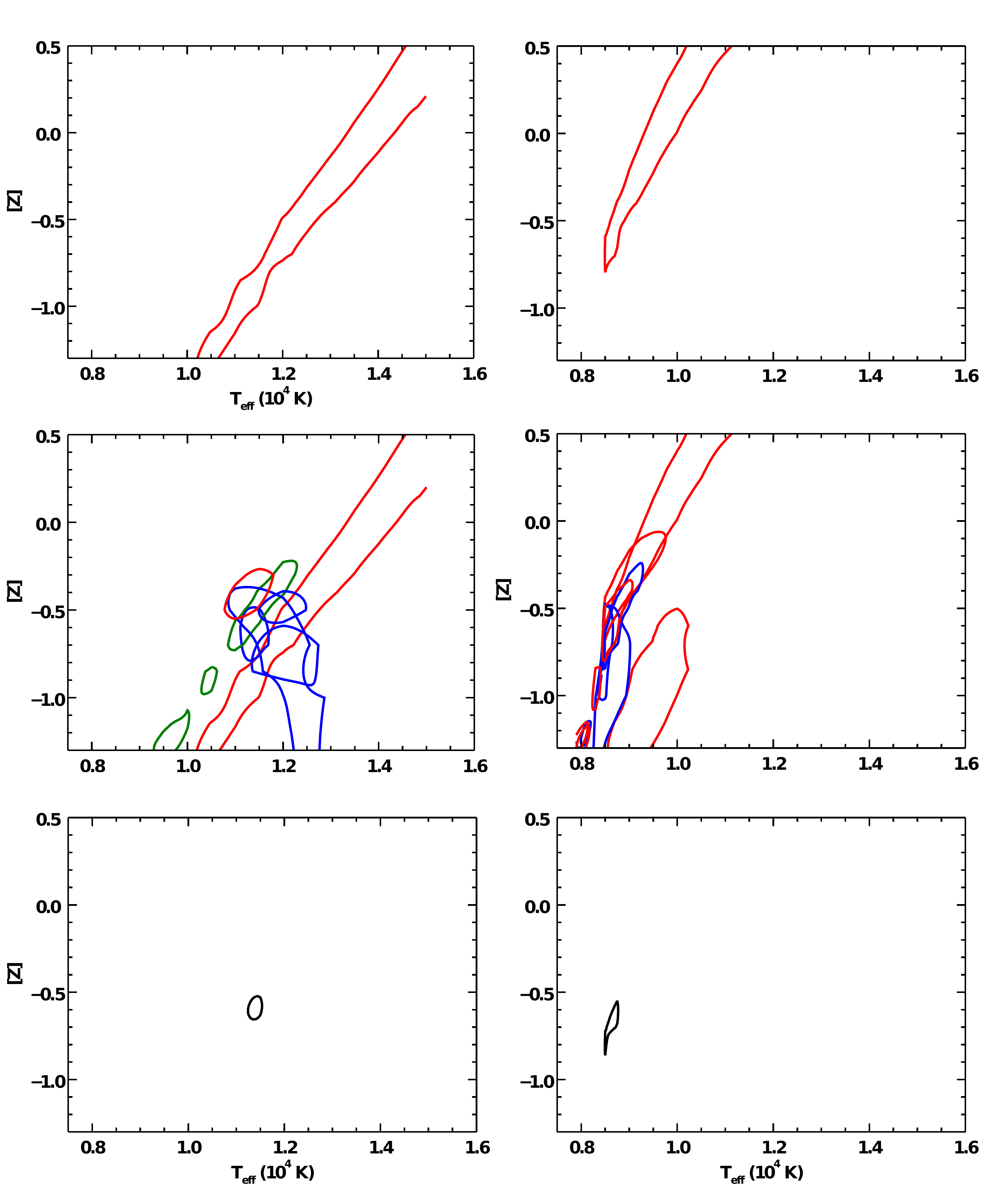}
\caption{A demonstration of our analysis method for a late-B (star 4, left) and early-A (star 17, right) star. \emph{Top:} 1$\sigma$ isocontour for a single spectral window (window 2). Note the temperature-metallicity degeneracy. \emph{Middle:} 1$\sigma$ isocontours of all the spectral windows. Blue isocontours are windows containing He I lines, while red isocontours represent windows without He I lines. For star 4, the window 1 isocontour is broken into three parts and is marked as green. Each isocontour line, though not restrictive on its own, provides important information about the stellar parameters. \emph{Bottom:} The resulting 1$\sigma$ isocontour when all of the spectral windows are added together. Note the remaining temperature-metallicity degeneracy in the early-A type star, due to the lack of constraining He I lines. See Appendix for the 1$\sigma$ and 2$\sigma$ isocontours of the remaining stars.}
\label{Analysis}
\end{center}
\end{figure}

With the spectral windows and continuum levels defined, we do a wavelength point-by-wavelength point comparison between the observed and model spectrum (with the model resolution degraded to match that of the observed spectrum) and calculate the $\chi^{2}$ statistic for each spectral window:
\begin{equation}
\chi^{2}(T_{eff}, [Z])  = \sum_{j=1}^{n_{w}} \frac{(F_j^{obs} - F_j^{model} )^2}{\sigma^2_s} 
\label{chieq}
\end{equation}
\begin{equation*}
\sigma_s = \frac{1}{(S/N)_s}
\end{equation*}
\\
where $n_{w}$ is the number of wavelength points for a given window, $(S/N)_s$ is the average signal-to-noise ratio across the spectrum, and F$_{j}^{obs}$ and F$_{j}^{model}$ are the normalized fluxes of the observed and model spectrum, respectively. We produce a grid of $\chi^2$ values for each spectral window across the full range of possible temperature and metallicity models, with the log \emph{g} of each model constrained by the Balmer fit. These individual windows show the temperature-metallicity degeneracy (Fig. \ref{Analysis}, top). However, the 1$\sigma$ isocontours of different windows cover different regions of the (T$_{eff}$, [Z]) plane (Fig. \ref{Analysis}, middle). By adding the $\chi^2$ values of the individual windows at each grid point, we combine this information to break the temperature-metallicity degeneracy. The He lines are especially valuable in this regard as their individual degeneracies are nearly perpendicular to those of the metal lines. All windows free of spectral defects are included in this analysis regardless of the position of their contours in the (T$_{eff}$, [Z]) plane, including those that do not appear to coincide with the other windows. This is done to safeguard against bias and to provide a conservative error estimate on the final stellar parameters. The overall T$_{eff}$ and [Z] of the object can then be found from the minimum in the combined $\chi^2$ grid  (Fig. \ref{Analysis}, bottom). To accurately determine the minimum and plot the surrounding $\Delta \chi^2$ isocontours (see below), we carry out a careful parabolic interpolation of the $\chi^2$ values at the model atmosphere grid points to a significantly finer grid. As discussed above, we do this in parallel for both the normal He abundance and enhanced He abundance models. 

\begin{figure}
\begin{center}
\includegraphics[scale=0.45]{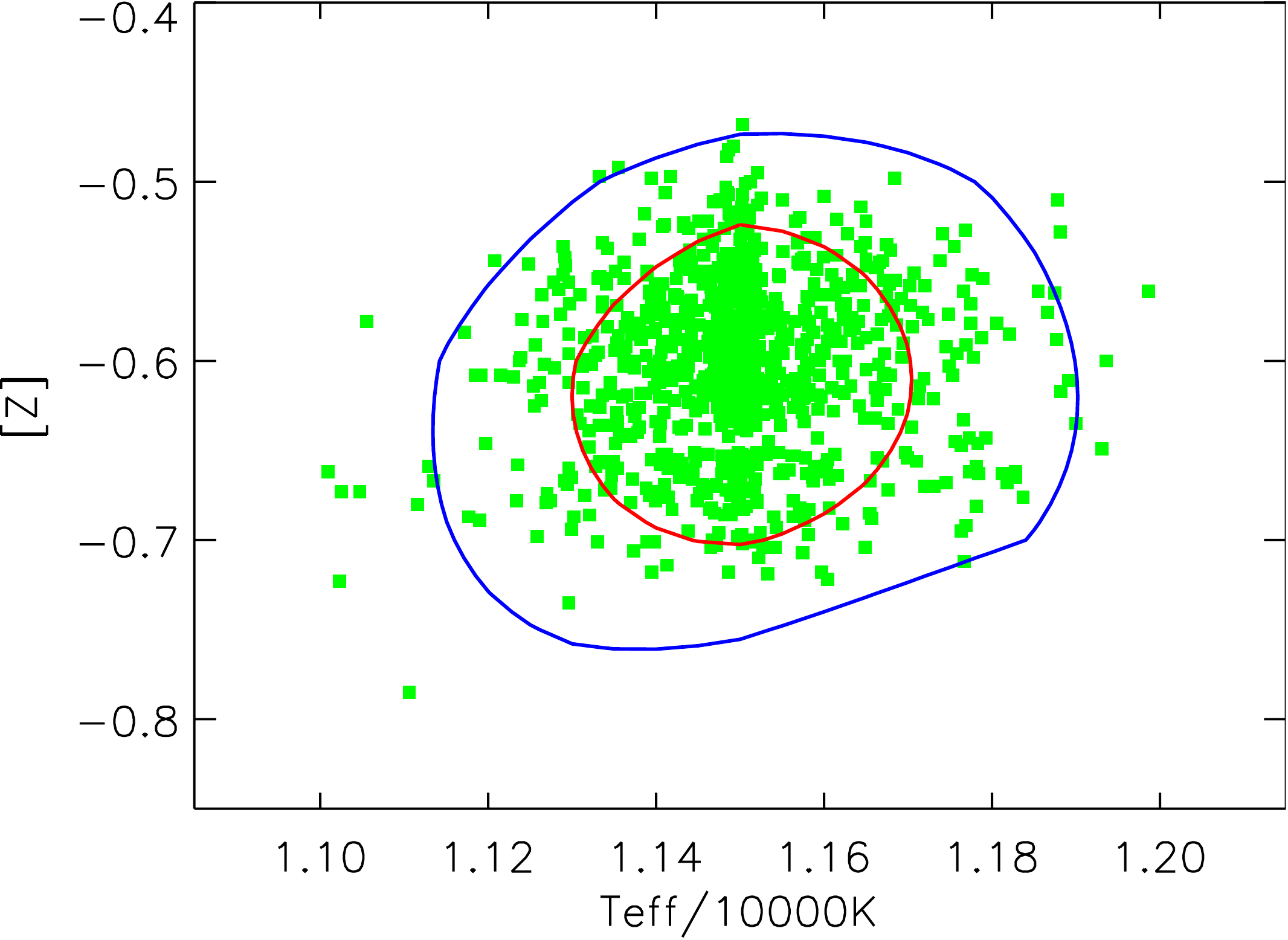}
\caption{An analysis of 1000 synthetic spectra with identical spectral parameters (T$_{eff}$ = 11500 K, [Z] = -0.6, log g = 1.70) but different gaussian noise, used to determine our 1$\sigma$ fit errors. The green markers indicate the extracted stellar parameters for each individual spectrum, while the red circle encompasses 68\% of the results and the blue circle encompasses 95\% of the results. These simulations show that 1$\sigma$ errors correspond to $\Delta\chi^2$~=~3.0 for our spectra.}
\label{simplot}
\end{center}
\end{figure}

To verify our method and to assess the $\chi^2$ uncertainties for our derived parameters, we analyze synthetic spectra with typical parameters for late-B and early-A BSGs (late-B: T$_{eff}$ = 11500 K, log \emph{g} = 1.75, [Z] = $-$0.7; early-A: T$_{eff}$ = 8750 K, log \emph{g} = 1.05, [Z] = $-$0.7). We generate two sets of 1000 individual spectra for both types, adding Gaussian Monte-Carlo noise to simulate S/N = 50 for one set and S/N = 100 for the other, and degrade the resolution to match that of our observed spectra. For each set, we run the individual spectra through our analysis, calculating a combined $\chi^2$ grid for each and determining the location of the minimum in the (T$_{eff}$, [Z]) plane. We also calculate a mean $\chi^2$ grid by averaging the 1000 individual $\chi^2$ grids and determine the location of the corresponding minimum $\overline{\chi^2_{min}}$. We then calculate [$\overline{\chi^2_{min}}$ + $\Delta\chi^2$] isocontours and identify the isocontours that encompass 68\%  and 95\% of the 1000 individual minimum locations (Fig. \ref{simplot}). In this way we establish the 1$\sigma$ and 2$\sigma$ $\chi^2$-fit uncertainties of our results \citep{Press07}. We find that $\Delta\chi^2$ = 3.0 corresponds to a conservative estimate of the 1$\sigma$ uncertainty. This is slightly higher than predicted by $\chi^2$ theory ($\Delta\chi^2$ = 2.3 when fitting 2 parameters), but this can be explained by uncertainties in the calculated spectral S/N and the fact that an average value is adopted over the entire spectral range, which affects the $\chi^2$ calculation (see Eq. \ref{chieq}). We also find that the $\Delta\chi^2$ -value is independent of the number of wavelength points and S/N, also in agreement with  $\chi^2$-theory \citep{Press07}.  The $\Delta\chi^2$ =  3.0  isocontour is then used to assess the 1$\sigma$ uncertainty in the stellar parameters of the observed spectra by taking the maximum and minimum parameter values within the isocontour. See appendix for the $\Delta\chi^2$ isocontours of the stars not presented in Fig. \ref{Analysis}.

The simulations show that our $\chi^2_{min}$-finding routine works well in recovering the parameters of both late-B and early-A type objects. The temperature-metallicity degeneracy for the late-B stars is broken completely, since these stars are hot enough for the He I lines to provide meaningful constraints. These lines are weak in the spectra of the cooler early-A stars, which as a result suffer from a partial degeneracy even after the combination of the spectral windows. This manifests itself as an elongated minimum ``valley'' in the combined $\chi^2$ grid (see Fig. \ref{Analysis} and appendix). 

There are several additional sources of error which must be taken into account for the derived stellar parameters. The effect of uncertainty in the continuum level is evaluated by performing a $\chi^2$ analysis using the highest and lowest possible values for each continuum region, calculated from the standard error of the median, and comparing the resulting stellar parameters to those determined in the original analysis. This uncertainty mainly affects [Z], and is added in quadrature with the 1$\sigma$ errors from the original $\chi^2$ fit to produce the final uncertainties reported. The total uncertainty in log \emph{g} is determined by combining the inherent uncertainty of the Balmer line fit with the change in the final log \emph{g} value caused by adopting the 1$\sigma$ maximum and minimum T$_{eff}$ values, which changes the log \emph{g} value according to the Balmer equivalent width isocontour. 

\subsection{Testing the Method}
As an independent test of our method, we analyzed the spectra of three SMC BSGs whose parameters are constrained by the high-resolution (R = 48000) and high-S/N (S/N $\approx$ 100) study by \citet{Sch10}. The resolution of the SMC spectra was degraded to match that of the NGC~3109 spectra and Gaussian Monte-Carlo noise was added to simulate S/N = 100. The results of this analysis are listed in Table \ref{SMC_results}. 

The comparison between our stellar parameters and those determined by Schiller (2010) is encouraging, as all of the results agree to within 1$\sigma$ (see fit isocontours in Appendix). Note that for the two hotter stars, AV76 and AV200, we derive a T$_{eff}$ with uncertainties smaller than what is reported in the high-resolution study. This is because the high-resolution temperature is determined from either the N I/II or Mg I/II ionization equilibrium, where the necessary lines are faint for hot stars at low metallicity. Since our low-resolution spectral synthesis method uses stronger lines from multiple elements we produce a competitive result. Though the high-resolution analysis indicates that all three stars have enhanced He, the best-fit normal and enhanced He abundance models give similar results. This is reassuring in the sense that the a priori adoption of a helium abundance does not significantly affect the resulting values of temperature, gravity and metallicity. The largest differences occur in the hotter stars, which is not surprising since high temperatures are required to produce He I lines strong enough to impact the analysis. 

\section{Results}

\subsection{Stellar Parameters}

\begin{figure}
\begin{center}
\includegraphics[scale=0.40]{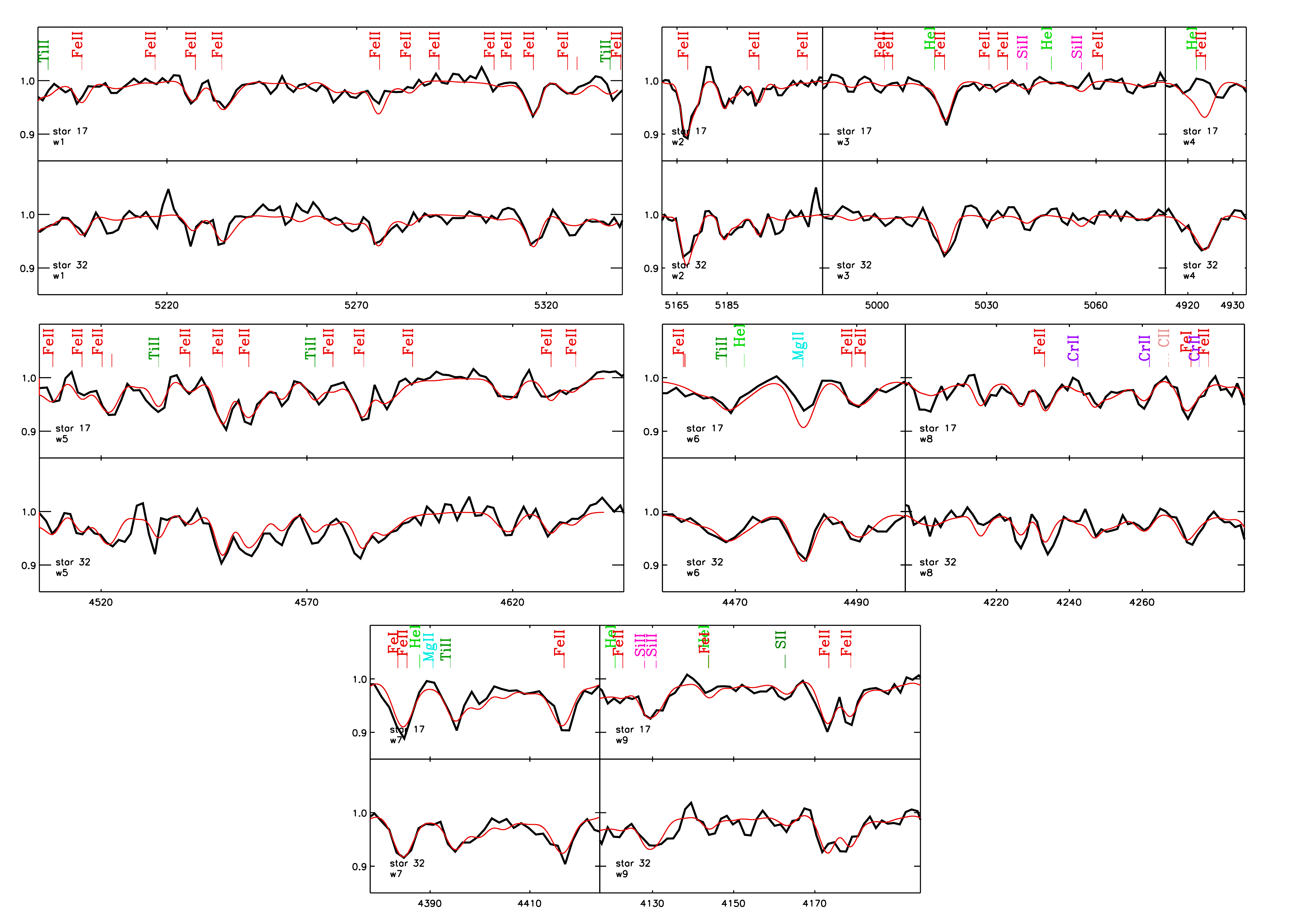}
\caption{A demonstration of the spectral fits from our method for stars 17 and 32, two early-A type stars. Observed spectrum is in black, best-model fit (normal He abundance) in red. The fits for the enhanced He abundance models are of similar quality.  The model fits of the other early-A type stars are provided in the appendix.}
\label{cool_metalfits}
\end{center}
\end{figure}

\begin{figure}
\begin{center}
\includegraphics[scale=0.40]{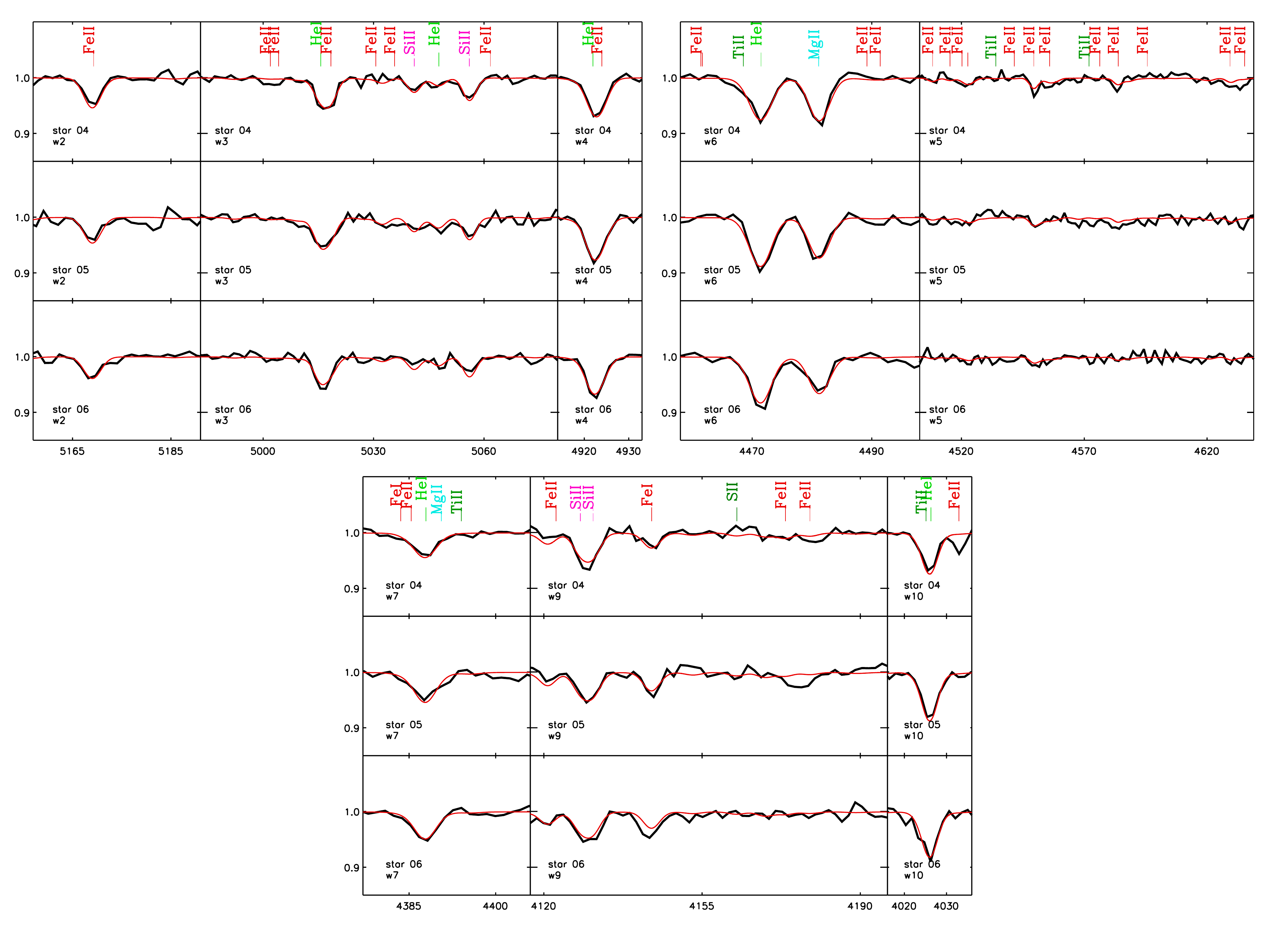}
\caption{A demonstration of the spectral fits from our method for stars 4, 5 and 6, three late-B type stars. Observed spectrum is in black, best-model fit (normal He abundance) in red. The fits for the enhanced He abundance models are of similar quality.  The model fits of the other late-B type stars are provided in the appendix.}
\label{hot_metalfits}
\end{center}
\end{figure}

Using our method, we successfully fit the observed spectra of our sample of late-B and early-A BSGs (see Figs. \ref{cool_metalfits}, \ref{hot_metalfits}, and Appendix). The stellar parameters we derive  are summarized in Table \ref{Stellarparams}, with the first entry for each star representing the best-fit normal He model and the second entry representing the best-fit enhanced He model. In addition to effective temperatures, gravities, and metallicities, we determine the total reddening E(B$-$V) (foreground + intrinsic) and bolometric correction BC for each star: E(B$-$V) by comparing the  observed V$-$I$_{c}$ color with the intrinsic V$-$I$_{c}$ color of the closest model to our best-fit parameters, and the BC from an analytical formula based on the model grid \citep[see][]{K08}. We then apply an extinction correction using the standard reddening law of R$_v$ = 3.1 \citep{Card89, OD94}. With these and the observed V magnitude (Table \ref{Myobjects}), the apparent bolometric magnitude m$_{bol}$ can be calculated. Errors for the bolometric magnitudes and stellar luminosities are obtained from the uncertainties of temperature, gravity, and metallicity, which affect the bolometric correction and the photometric uncertainty of reddening. Typical errors amount to $\sim$0.3 mag, dominated by uncertainties in the effective temperature.  At the level of reddening encountered in NGC~3109, the uncertainty produced by possible deviations from the standard reddening law is small compared with these errors.

In the framework of our analysis, we cannot determine which He abundance (normal or enhanced) provides a better fit to our observed spectra, and so we present the results of both model grids in parallel. However, an inspection of Table \ref{Stellarparams} reveals that there is very little, if any, difference between the parameters of the early-A stars, and that the choice of He abundance only affects the hotter late-B stars. This is consistent with the test analysis of the SMC spectra. Table \ref{Chimin_comp} provides the reduced chi-squared $\chi^2_{\nu}$ = $\frac{\chi^2_{min}}{\nu}$, where $\nu$ is the degrees of freedom and is approximately equal to the total number of pixels in all of the spectral windows, of our spectral fits. Theory predicts $\chi^2_{\nu} \approx 1$ for a good fit between the model and observations \citep{Bev03}, though in our case uncertainties in the spectral S/N prevent a rigorous statistical interpretation of $\chi^2_{\nu}$. We include the reduced chi-squared to give the reader a general sense of our fits. As we discuss below, the choice of He abundance has little effect on our overall results for NGC 3109.

\begin{figure}
\begin{center}
\includegraphics[scale=0.35]{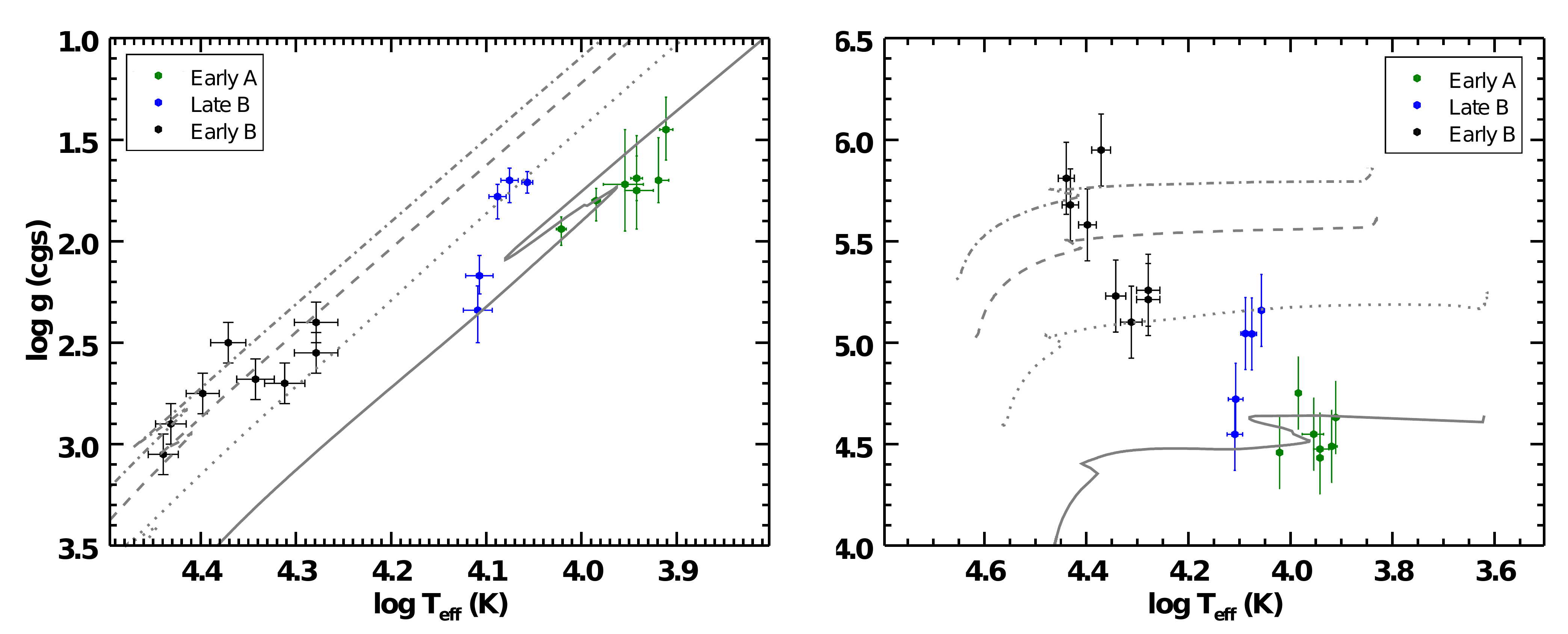}
\caption{Comparing the sample of BSGs to evolutionary tracks in the (log \emph{g} vs. T$_{eff}$) plane (left) and H-R diagram (right). Stellar parameters are from the best-fit normal He abundance models, though the choice of He abundance does not have a significant impact. The apparent grouping of stars within a given spectral type is caused by differences in mass. The evolutionary tracks correspond to initial masses of 12, 20, 30, and 40 M$_{\odot}$, assuming [Z] = $-$0.7 and an initial rotation velocity of 300 km~s$^{-1}$ \citep{MM01, MM05}. }
\label{Ev_tracks}
\end{center}
\end{figure}

We combine our late-B and early-A objects with the early-B objects analyzed by E07 (Table \ref{Stellarparams_E07}) to build a total sample of 20 BSGs in NGC~3109. We compare the position of each object in the (log \emph{g}, log $T_{eff}$) plane and H-R diagram with the low-metallicity evolutionary tracks by \citet{MM01} and \citet{MM05}. These tracks are calculated for 12-40~M$_{\odot}$ stars at [Z] = -0.7, and incorporate the effects of stellar rotation. The advantage of the (log \emph{g}, log $T_{eff}$) diagram is that it only relies on spectroscopic parameters; the H-R diagram, on the other hand, requires adopting the distance modulus we find using the FGLR ($\mu$ = 25.55, see $\mathsection$ 3.2). Both diagrams are consistent with each other, showing the expected progression between different spectral types (Fig. \ref{Ev_tracks}). The evolutionary tracks indicate that the individual stellar masses lie approximately between 12-40 M$_{\odot}$, as expected for BSGs. This is an encouraging check of our stellar parameters. We also note that contrary to the studies of M81 and M33 by \citet{K12} and \citet{U09}, respectively, we do not find objects for which the location in the (log \emph{g}, log $T_{eff}$) diagram indicates a mass significantly different from the one in the HRD.

\begin{figure}
\begin{center}
\includegraphics[scale=0.55]{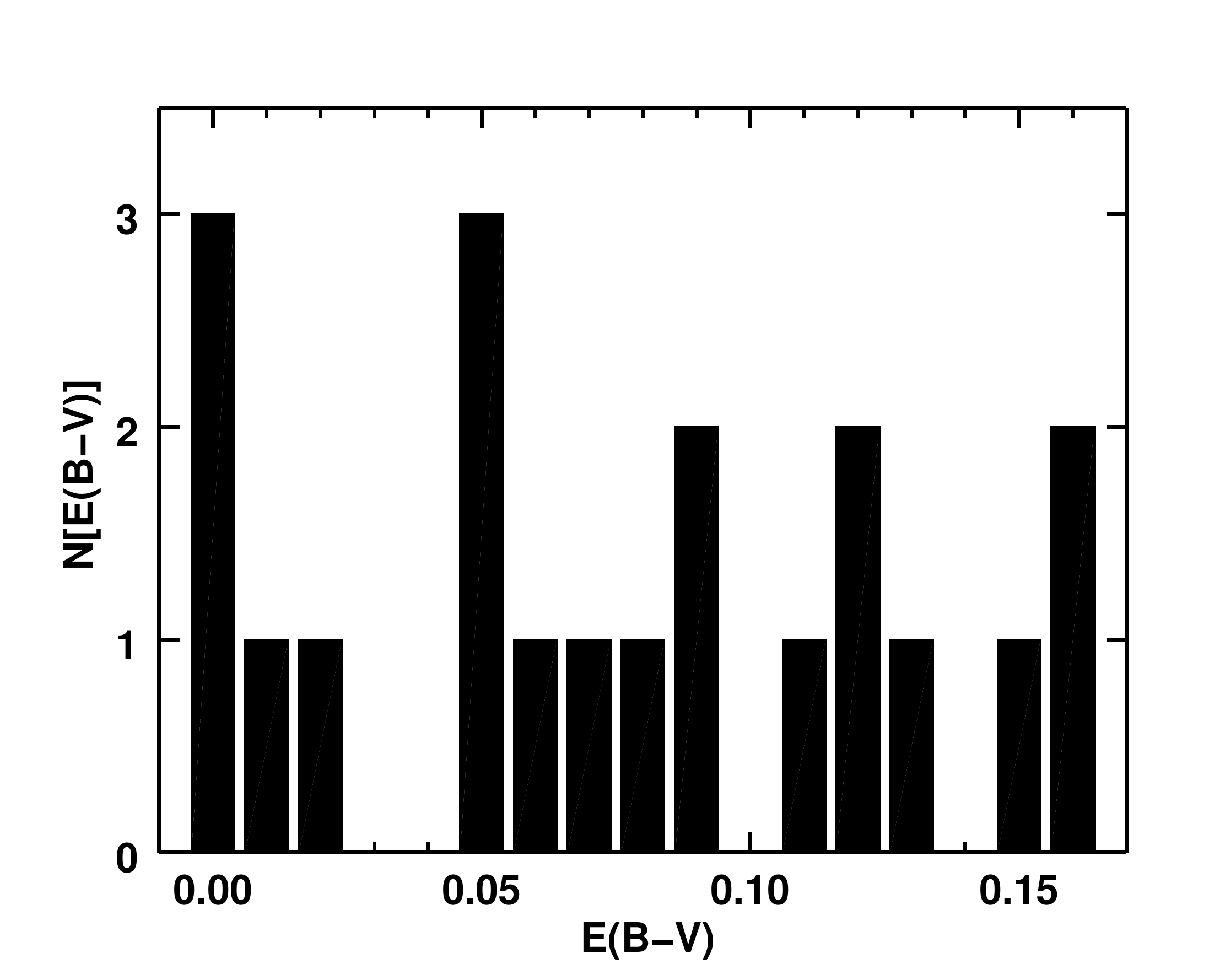}
\caption{Reddening values of the combined BSG sample. E(B$-$V) accounts for both foreground and intrinsic reddening, and has a typical uncertainty of 0.03 mag. Reddening values associated with the normal He abundance models are shown, though the choice of He abundance does not have a significant effect.}
\label{Reddening}
\end{center}
\end{figure}

We find a range of reddening values for our combined sample of BSGs (Fig. \ref{Reddening}; 0.0 $\le$ E(B$-$V) $\le$ 0.16), indicating that these objects have varying amounts of intrinsic reddening. For the early-B BSGs we re-determine the reddening using a more recent version of the FASTWIND model atmosphere code \citep{SR97,Pu05} than was used by E07 for their analysis. The new values of E(B-V) are very similar to E07. However, for stars 7 and 22, where we find slightly negative reddening similar to E07, we adopt E(B-V)=0.0 mag. This is different from E07, who assign an average reddening value of 0.09 mag to these objects. The typical uncertainty in E(B$-$V) is $\sim$0.03 mag. Given these uncertainties, our reddening values are consistent with the galactic foreground reddening of E(B$-$V) = 0.06 $\pm$ 0.02 \citep{G11, Sc98} in the direction of NGC~3109, with several objects showing additional intrinsic reddening. The average reddening of E(B$-$V) = 0.07 is similar to the Cepheid variable studies of \citet{So06} and Pietrzy\'{n}ski et al. (2006), who find E(B$-$V) = 0.087 $\pm$ 0.012 and E(B$-$V) = 0.10, respectively. Conflicting claims have been made regarding the presence of differential reddening within NGC~3109; \citet{M99} suggest that the east side has $\sim$0.1 mag more extinction than the west side, while \citet{H08} do not find such a discrepancy. We do not see evidence of a large extinction difference across the galaxy, though our sample is too small to discount the possibility. 

\subsubsection{Spectroscopic vs. Evolutionary Mass}
From the luminosity, two measures of stellar mass can be calculated: spectroscopic mass (M$_{spec}$) using the stellar radius and gravity derived from the spectrum, and evolutionary mass (M$_{evol}$) using the BSG mass-luminosity relationship derived by \citet{K08} from evolutionary tracks with SMC metallicity (for the determination and discussion of metallicity, see the following subsection and subsection 4.3.2). A comparison of these masses provides an additional test of our results, identifying stars perhaps affected by binary star or blue loop evolution (e.g. \citeauthor{K08} 2008; \citeauthor{U09} 2009), and, in our case, a confirmation of the parameters for early-A stars still affected by a weak temperature-metallicity degeneracy. It also offers a method of examining the systematics between evolutionary tracks and model atmospheres. Past studies have found that the spectroscopic masses of massive stars are often systematically lower than their evolutionary masses (e.g. \citeauthor{He92} 1992), though recent studies with improved model atmospheres have shown that this effect has been significantly reduced \citep{K08, U08, U09, K12}. 

\begin{figure}
\begin{center}
\includegraphics[scale=0.35]{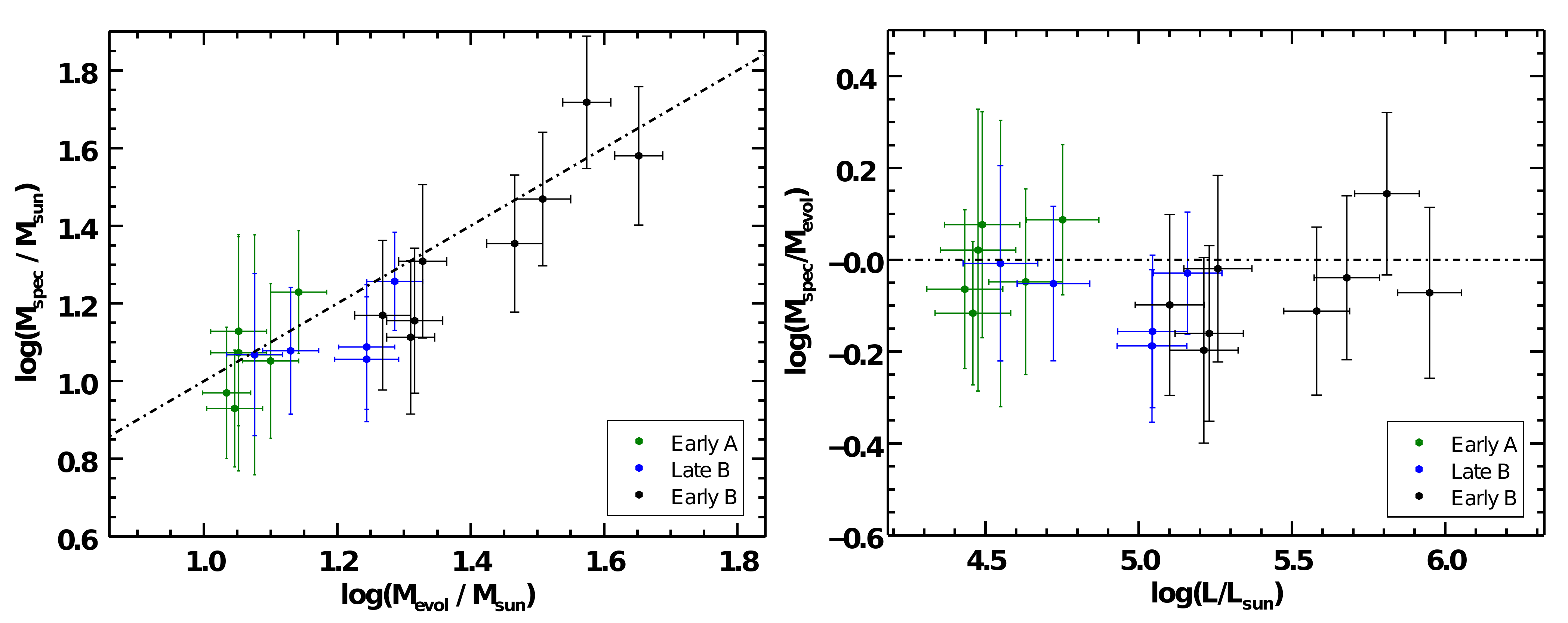}
\caption{\emph{Left:} A comparison of spectroscopic and evolutionary masses. While the spectroscopic mass is on average 0.05 dex less than the evolutionary mass, the values generally agree to within uncertainties. \emph{Right:} The logarithmic ratio of spectroscopic to evolutionary mass as a function of luminosity. No systematic trend is observed.}
\label{Masscomp}
\end{center}
\end{figure}

The absolute bolometric magnitudes, luminosities, radii, and spectroscopic/evolutionary masses calculated for each object are presented in Table \ref{Masscomp_table}. The spectroscopic masses are more uncertain than the evolutionary masses because they incorporate uncertainties in both T$_{eff}$ and log \emph{g}, though the evolutionary masses are prone to systematic errors in the evolutionary tracks. A comparison of the masses shows that while M$_{spec}$ is typically $\sim$0.05 dex lower than M$_{evol}$, they generally agree within uncertainties (Fig. \ref{Masscomp}). No clear trend is found between the mass discrepancy and luminosity, and no obvious outliers exist. This indicates that single star evolution can explain the objects well and that additional mass-loss processes caused by blue-loop evolution back from the red supergiant branch or binary mass exchange are not important. This is in good agreement with the previous studies of BSGs analyzed using this grid of model atmospheres and serves as an affirmation of our derived stellar parameters. 

\subsubsection{Metallicity}
\citet{K08} showed that individual BSGs can be used as reliable metallicity indicators within galaxies based on the many metal lines in their spectra, which vary strongly as a function of metallicity (see their Fig. 11). The reliability of this method was further demonstrated by \citet{B09}, who found the BSG metallicities and subsequent abundance gradient derived for NGC 300 by \citet{K08} to be highly consistent with those determined from HII regions via direct measurement using auroral lines. Moreover, they found that applying several strong-line calibrations to their HII region sample produced significantly different metallicities, emphasizing the importance of BSGs when studying galaxy metallicities where auroral line measurements are not possible. 

\begin{figure}
\begin{center}
\includegraphics[scale=0.35]{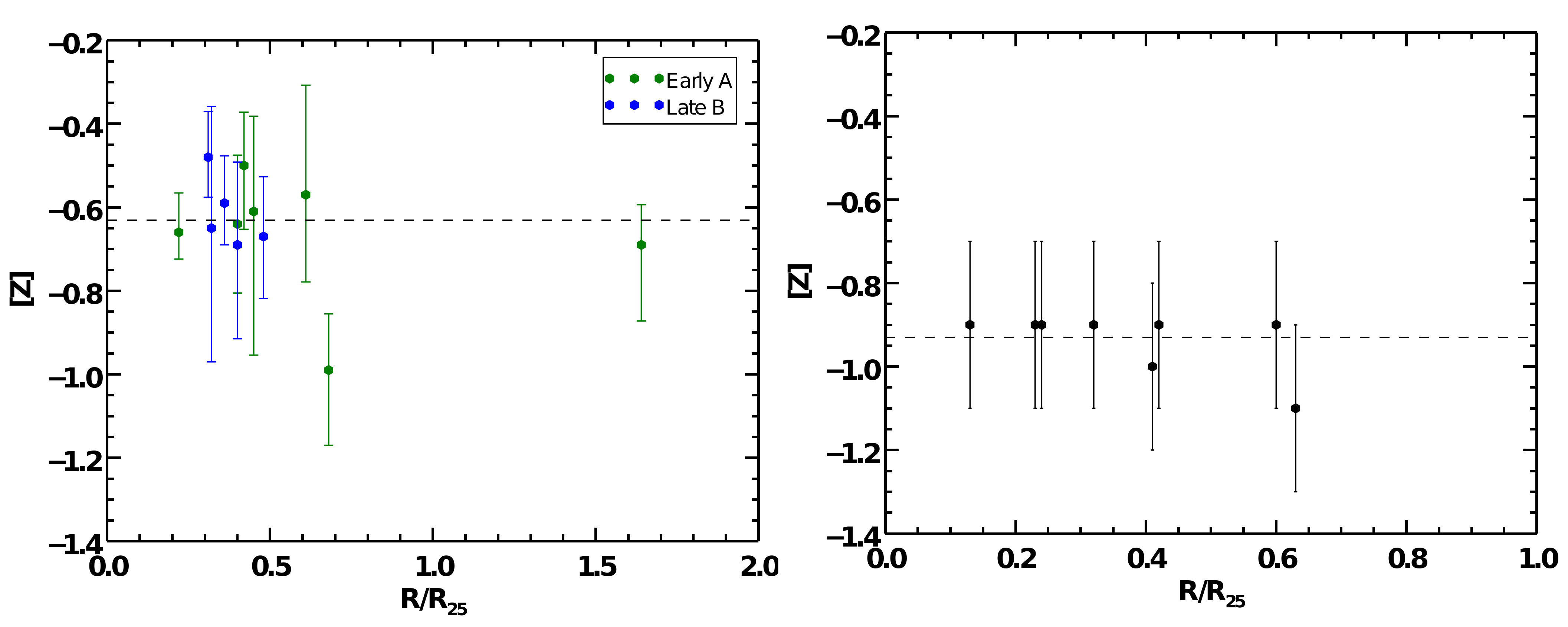}
\caption{\emph{Left:} Metallicity vs. de-projected galactocentric distance for the late-B and early-A objects, using normal He abundance models. The dotted line shows the weighted average of the sample. \emph{Right:} Same as left, only for the early-B objects analyzed by \citet{E07}. Note the discrepancy in metallicity between the samples, which were obtained using different elements. No suggestion of a metallicity gradient is found.}
\label{Mgrad}
\end{center}
\end{figure}

We determine an average metallicity for NGC~3109 from the weighted average of the metallicities of our late-B and early-A type objects: 

\begin{equation}
[\bar{Z}] = \frac{\sum_{i = 1}^{n_{s}}w_i [Z]_i}{\sum_{j = 1}^{n_{s}}w_j}
\end{equation}
\begin{equation*}
w_i = \frac{1}{\sigma^2_i}
\end{equation*}
\\
where $n_{s}$ is the number of late-B and early-A stars in our sample while $[Z]_i$ and $\sigma_i$ are the metallicity and uncertainty in metallicity for a given star, respectively. The choice of He abundance has a small effect on this result due to the shift in metallicity for the late-B stars, which is driven by a slight change in $T_{eff}$. Because the strength of the He I lines increases with both $T_{eff}$ and He abundance, the enhanced He model fits have slightly lower temperatures than the normal He abundance model fits. Since the metal line strengths increase with decreased temperature, lower metallicities can be used to fit the observed metal lines, and thus the enhanced He abundance models have lower metallicities than the normal He abundance models. This effect is not seen with the early-A stars since they do not have strong He I lines. If the normal He abundance models are used then we find [\={Z}]~=~$-$0.63 $\pm$ 0.13, while if the enhanced He abundance models are used we find [\={Z}]~=~$-$0.71 $\pm$ 0.14. The uncertainty on these values are calculated from the standard deviation of the sample. In both He abundance cases, all but two of the late-B and early-A stars (stars 5 and 25) agree with the associated average metallicity to within one sigma of their error. Interestingly, both of these average metallicities are greater than one sigma above than the  value of [\={Z}] = $-$0.93 $\pm$ 0.07 found by E07 from the early-B objects (Fig. \ref{Mgrad}) and supported by HII region studies \citep{Le03, Le03b, P07}. It is important to note that our metallicity is derived from Fe-group elements, while the E07 result is derived from the oxygen. We discuss this discrepancy in $\mathsection$ 4.3.2. Moving forward, we adopt a final metallicity of  [\={Z}]~=~$-$0.67 $\pm$ 0.13 for NGC 3109, the average of the normal and enhanced He abundance metallicities. 

Given that our late-B and early-A objects are fairly spread out along the galactic disk, we can use our results to test for a metallicity gradient in NGC~3109. The presence of such a gradient would indicate a changing star formation history with galactocentric distance, as commonly found in spiral galaxies (see references in $\mathsection$ 1). Past spectroscopic studies of BSGs have been used to investigate metallicity gradients in NGC 300 (Kudritzki et al. 2008), M33 (U et al. 2009), M81 (Kudritzki et al. 2012), and NGC 55 \citep{C12}. We adopt the positional parameters of Jobin \& Carignan (1990, Table \ref{3109_pos}) to calculate the de-projected galactocentric distance of each star. These distances should be treated cautiously, as the high inclination angle of NGC~3109 makes such calculations uncertain. Even so, we do not find evidence of a significant abundance gradient in NGC~3109 in either the normal or enhanced He abundance case (Fig. \ref{Mgrad}). This is consistent with the low dispersion in oxygen abundances found in early-B BSGs and HII regions (Evans et al. 2007; Pe\~{n}a et al. 2007) and indicates that NGC~3109 is fairly homogenous in terms of metallicity and star formation history.

\subsection{FGLR and Distance Modulus}
As discussed in detail by \citet{K03, K08}, the fact that BSGs evolve at roughly constant mass and luminosity leads to a tight correlation between the flux weighted gravity g$_F$ (g$_F \equiv$ g/T$^{4}_{eff}$, where T$_{eff}$ is in units of 10$^4$ K and g is in cgs units) and the absolute bolometric magnitude M$_{bol}$. Called the Flux-weighted Gravity-Luminosity Relation (FGLR), this can be used to calculate galactic distances independent of Cepheid variables, tip of the red giant branch (TRGB), and other methods. The advantages of the FGLR and its previous applications are summarized in $\mathsection$ 1. 

\begin{figure}
\begin{center}
\includegraphics[scale=0.55]{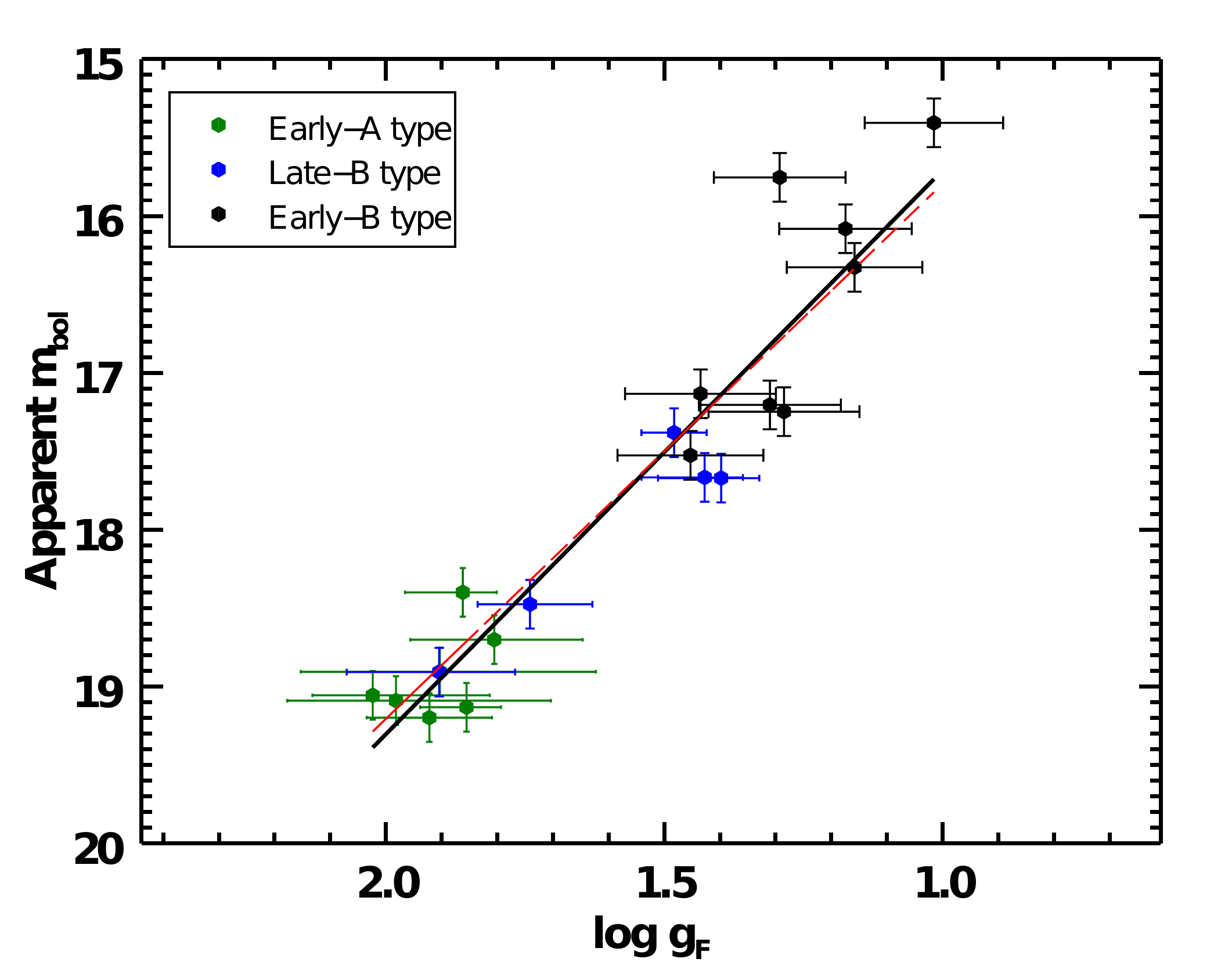}
\caption{The observed FGLR for NGC~3109. Note that the early-B stars from E07 have been merged with the late-B and early-A stars from this study. The black line is a fit to the data alone, while the dashed red line is the fit after fixing the slope to that of the calibrated FGLR. The difference between the two fits is negligible.}
\label{obsFGLR}
\end{center}
\end{figure}

The FGLR has the form:

\begin{equation}
M_{bol} = a(log~g_{F} - 1.5) + b
\end{equation}
\\
where a = 3.41 and b = $-$8.02 using the calibration of \citet{K08}. Calculating log g$_F$ for the late-B and early-A BSGs, as well as the early-B BSGs from E07, and plotting against apparent m$_{bol}$ reveals an observed relation of

\begin{equation}
m_{bol} = a_{3109}(log~g_F - 1.5) + b_{3109}
\end{equation} 
\\
where $a_{3109}$ = 3.59 $\pm$ 0.11 and $b_{3109}$ = 17.51 $\pm$ 0.04 in the normal He abundance case as shown in Fig. \ref{obsFGLR}. Very similar values are obtained for the enhanced He abundance case, which is why we only discuss the normal He abundance values here. We note that the spectroscopic determination of log \emph{g} and T$_{eff}$ for B- and A-type supergiants is only very weakly influenced by [Z], and so the metallicity offset between our sample and the E07 sample will not have a significant effect on the FGLR. This is also why our choice of He abundance does not change our FGLR results, since the best-fit models primarily differ in metallicity, if at all. 

By fixing the slope of the observed relation to the calibrated slope of the FGLR ($a_{3109} = a$), we can fit for only $b_{3109}$. Due to the agreement between the slopes, the value we recover for $b_{3109}$ is nearly identical to that found from the pure fit of the data. The difference between $b$ and $b_{3109}$ then yields the distance modulus $\mu$. We find $\mu$ = 25.54 $\pm$ 0.09 for the normal He abundance models, and $\mu$ = 25.56 $\pm$ 0.09 for the enhanced He abundance models. The error is calculated from the variance of the fit \citep{Bev03}:

\begin{equation}
\sigma^2_\mu = \frac{s^2_{fit}}{N} = \frac{1}{N(N-1)}  \sum^N_{i=1} w_i(M_{bol, i} - M_{bol, i}^{FGLR})^2
\end{equation}
\begin{equation*}
w_i = \frac{1/\sigma^2_i}{(1/N) * \sum_{j=1}^N(1/\sigma^2_j)}
\end{equation*}
\\
where $M_{bol}$ is the absolute bolometric magnitude calculated from photometry and the distance modulus, $M_{bol}^{FGLR}$ is the absolute bolometric magnitude calculated from the FGLR, $\sigma$ is the error in $M_{bol}^{FGLR}$ calculated for each star, and N is the total number of objects. Our results show that the distance modulus is robust regardless of which He abundance is adopted, as discussed above, and so we adopt a final distance modulus of $\mu$ = 25.55 $\pm$ 0.09 (1.27 Mpc) for NGC 3109.

\begin{figure}
\begin{center}
\includegraphics[scale=0.7]{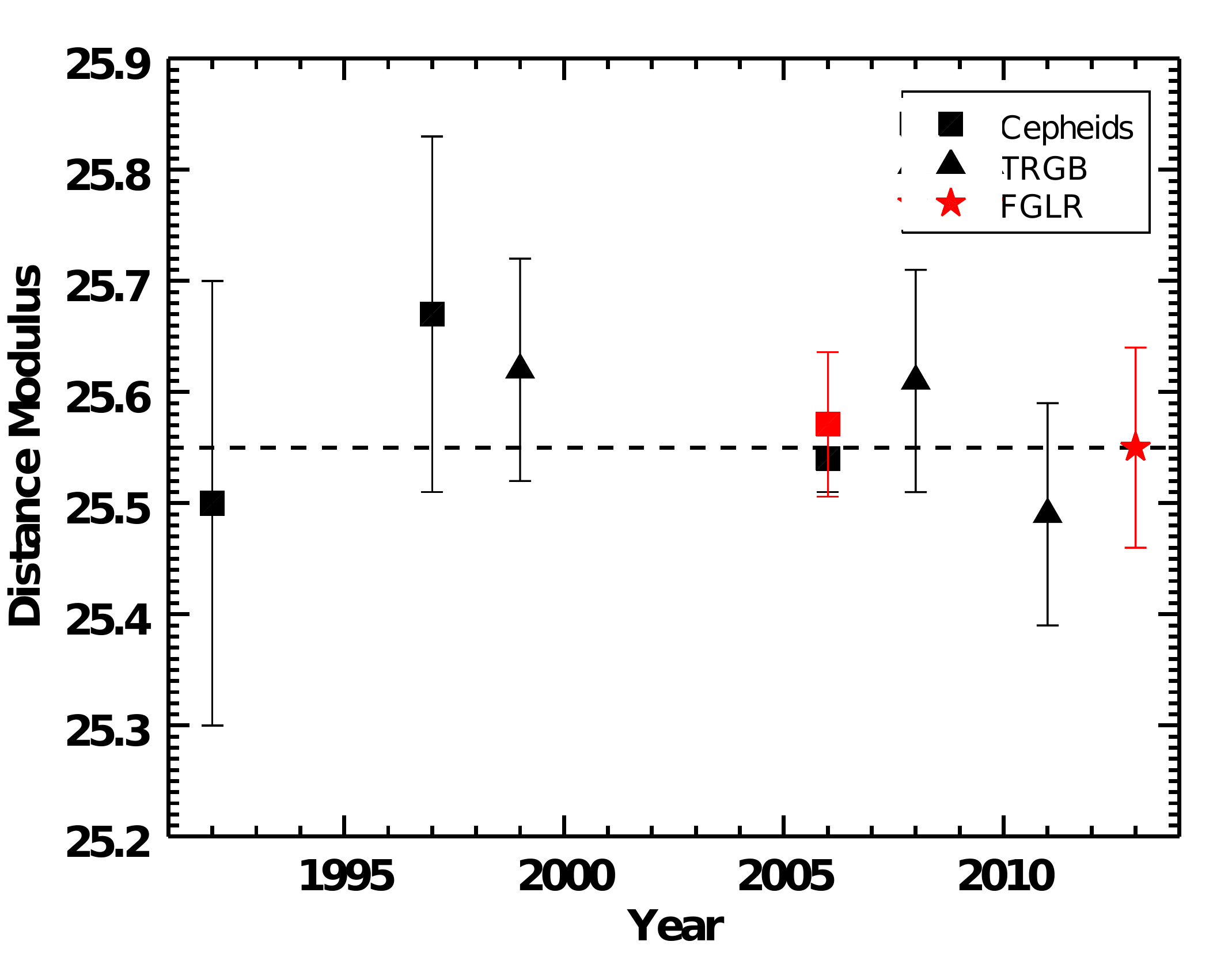}
\caption{The distance modulus of NGC~3109, as reported in the literature (references in text). Squares and triangles correspond to the Cepheid and TRGB distances, respectively. The red square is IR-based cepheid study of \citet{So06}, which is least affected by reddening. The red star and dotted line show the distance modulus found in this study using the FGLR. }
\label{Distcomp}
\end{center}
\end{figure}

Our FGLR distance modulus agrees well with values derived in the recent literature, as summarized by Fig. \ref{Distcomp}. Optical studies of Cepheids in NGC~3109 have yielded  $\mu$ = 25.5 $\pm$ 0.2 \citep{Ca92}, 25.67 $\pm$ 0.16 \citep{Mu97}, and 25.54 $\pm$ 0.03 (Pietrzy\'{n}ski et al. 2006), while IR photometry by Soszy\'{n}ski et al. (2006) found $\mu$ = 25.571 $\pm$ 0.024 (stat) $\pm$ 0.06 (syst). We note that the IR determined distance is likely the most reliable, since it is least affected by reddening and metallicity. NGC~3109 has also been the target of several TRGB studies, which have found $\mu$ = 25.62 $\pm$ 0.1 (Minniti et al. 1999), 25.61 $\pm$ 0.1 (Hidalgo et al. 2008), and 25.49 $\pm$ 0.05 (stat) $\pm$ 0.09 (syst) by G\'{o}rski et al. (2011). All of these values agree within the margin of their uncertainties to our distance.

\section{Discussion}
The success of our analysis method in producing stellar parameters and an FGLR for BSGs from NGC 3109 demonstrates that the Balmer jump does not need to be relied upon for quantitative low-resolution spectroscopy of late-B and early-A type stars. There are two main disadvantages of this approach, however. First is the need to adopt a He abundance for the model atmospheres, which primarily affects the late-B type stars whose temperatures are high enough for the He I lines to play a role in the analysis. Since BSGs are frequently found to have enhanced He abundances, these models must be taken into careful consideration. While the He abundance does not have a significant effect on the temperatures and gravities (and thus the FGLR) of our sample, it does lead to small changes in metallicity. The second disadvantage is that the temperature-metallicity degeneracy of the early-A type BSGs cannot be completely broken, since the He I lines are not strong enough to constrain the temperature as effectively as for the late-B stars. As a result, these stars have larger uncertainties in their parameters than late-B type stars with spectra of similar quality would. That said, our method generally produces stellar parameters with errors similar to those found in studies using the Balmer jump for spectra with S/N $\ge$ $\sim$50. This is a valuable simplification of the BSG analysis method, since large ground-based telescopes with near-UV sensitive instruments are rare, and accurate ground-based near-UV flux calibrations are always challenging. 

\subsection{Consistency with Cepheid Distances: Evidence That the Cepheid PL Relation Is Not Significantly Affected by Metallicity}

It is not yet settled how metallicity affects the Cepheid period-luminosity (PL) relation, and so the application (or lack thereof) of a metallicity correction remains a potential source of systematic error in Cepheid-based distances \citep[see discussion in][]{K12,Maj11,St11}. This question has been pursued by the ongoing Araucaria Project, which strives to precisely measure the distances to nearby galaxies in an effort to better understand how different distance determinations are affected by environmental factors \citep{Pi02,G05}. It is under the Araucaria Project that the Cepheid studies of NGC~3109 by \citet{Pi06} and \citet{So06} were undertaken. These studies find the observed optical (V, I) and IR (J, K) PL slopes to be consistent with those observed in the higher-metallicity LMC ([Z] = $-$0.3), suggesting that metallicity does not have a significant effect between $-$0.3 \textless [Z] \textless $\sim$$-$0.7. As a result, they calculate their distances by fixing the NGC~3109 PL slope to the LMC slopes. That our FGLR distance is consistent with their results strengthens this conclusion; if NGC~3109 had a different PL relation due to its low metallicity, then our independently-determined distance would not agree. This is in concurrence with other investigations by the Araucaria Project which have indicated that the Cepheid PL relation is largely independent of metallicity between $-$0.3 \textless [Z] \textless $-$1.0 \citep{Pi06_IC}.

\subsection{The FGLR of NGC~3109: Comparison With Other Galaxies}

\begin{figure}
\begin{center}
\includegraphics[scale=0.6]{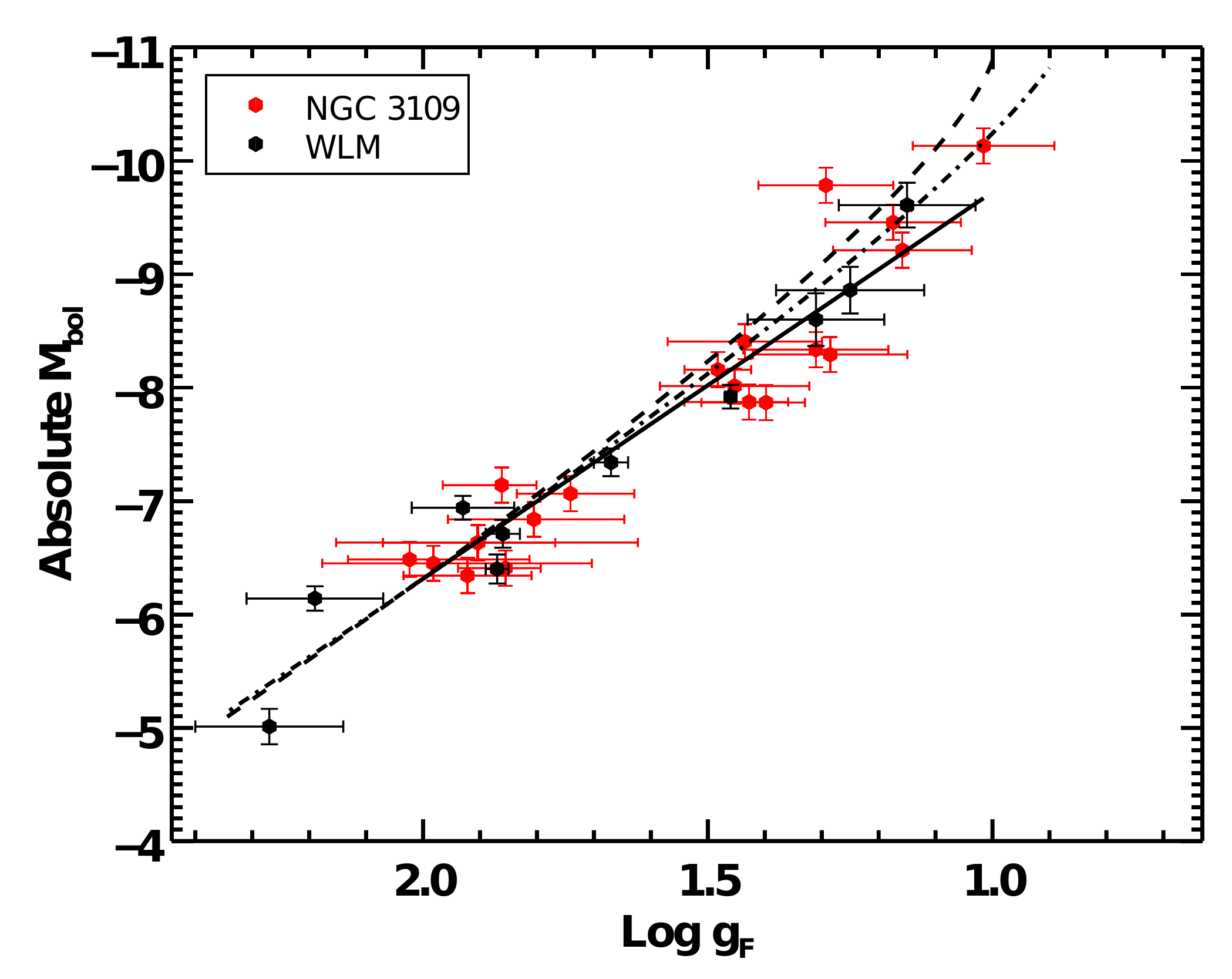}
\caption{The combined FGLR of NGC~3109 and WLM, both low-metallicity galaxies ([Z] \textless $-$0.6). Also plotted are the theoretical FGLRs for SMC metallicity (dashed line) and solar metallicity (dash-dotted line) of \citet{K08} based on the evolutionary tracks of \citet{MM05}. The calibrated FGLR (black line) of \citet{K08} is included for comparison.}
\label{LowZtheory}
\end{center}
\end{figure}

While the late-B and early-A BSGs indicate that NGC~3109 is not as quite as metal-poor as found by previous studies, the galaxy still provides an opportunity to investigate the behavior of the FGLR at low metallicities. In Fig. \ref{LowZtheory} we plot the NGC~3109 BSGs (early-B stars of E07 included) with a sample of BSGs in WLM, another low-metallicity galaxy ([\={Z}] = $-$0.87) analyzed by Urbaneja et al. (2008). Included for comparison are the FGLRs predicted by stellar evolution for SMC ([\={Z}] = $-$0.7) and  solar metallicities (Meynet \& Maeder 2005; Kudritzki et al. 2008), which incorporate the effects of stellar rotation with an adopted initial rotation velocity of 300 km s$^{-1}$. While the theoretical FGLRs are linear and nearly identical at high log $g_F$, they begin to show curvature at low log $g_F$ with the SMC-metallicity FGLR curving stronger toward higher luminosities than the solar metallicity one (see discussion in Kudritzki et al. 2008). This general behavior appears to be supported by the FGLRs for NGC~3109 and WLM, as  the higher-luminosity objects (primarily the early-B BSGs) begin to deviate from a linear track in a similar manner. The present uncertainties in log $g_F$ and M$_{bol}$ as well as the lack of more objects at the high luminosity end prevents further interpretation at this point.

\begin{figure}
\begin{center}
\includegraphics[scale=0.6]{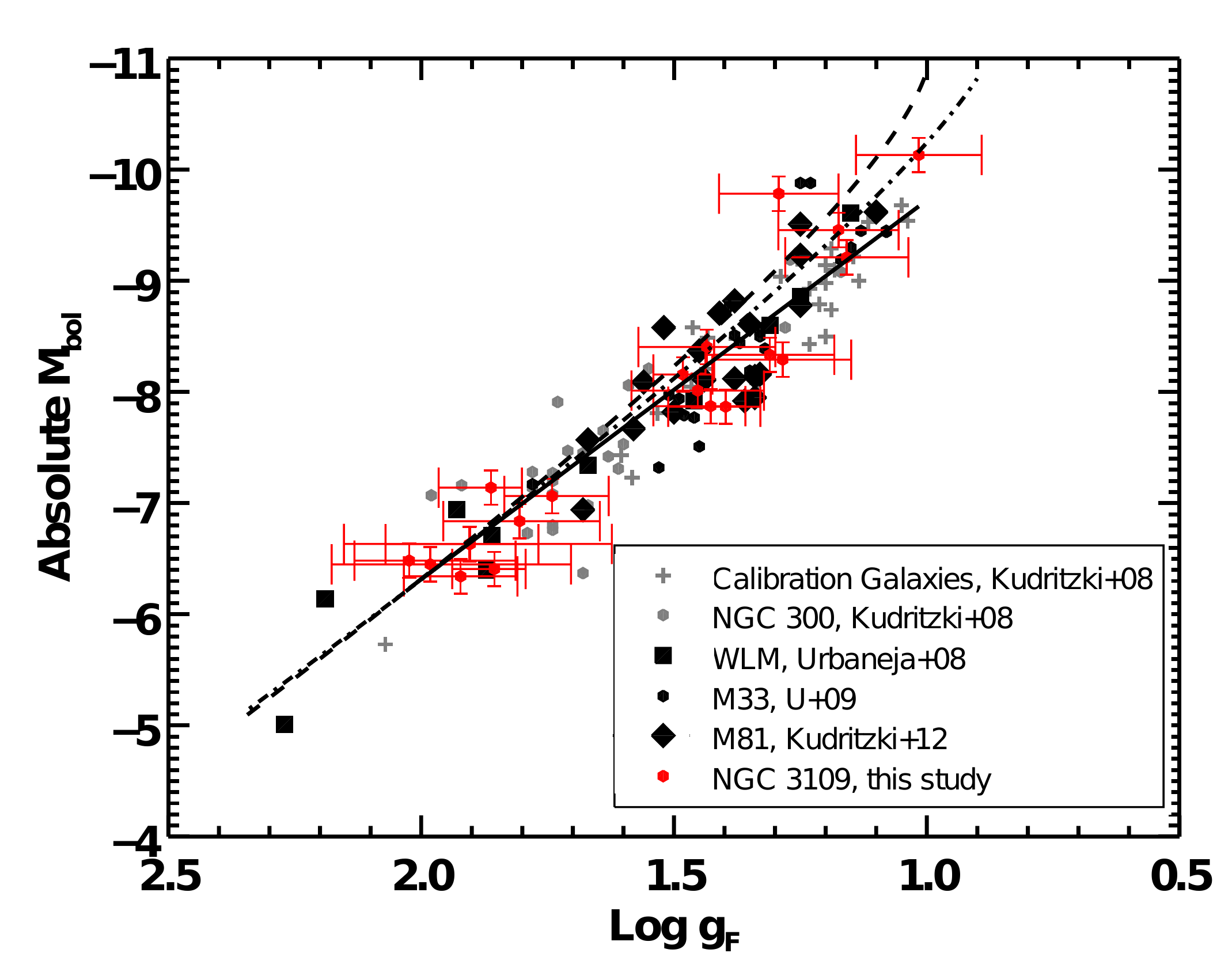}
\caption{The combined FGLR of all galaxies studied thus far. The references are as follows: NGC~3109 (this study, E07), M81 \citep{K12}, M33 \citep{U09}, WLM \citep{U08}, NGC 300 \citep{K08} and calibration galaxies \citep{K08}. The calibrated FGLR of \citet{K08} is represented by the black line, with the theoretical FGLRs for SMC and solar metallicity by the dashed and dash-dotted lines, respectively \citep{K08, MM05}.}
\label{AllFGLR}
\end{center}
\end{figure}

The FGLR of NGC~3109 is largely consistent with the FGLRs measured in other galaxies (Fig. \ref{AllFGLR}), a further validation of our analysis method. On the whole, the consistency of the FGLRs is highly encouraging given the wide range of flux-weighted gravities (1.02 $\ge$ log $g_F$ $\ge$ 2.27) and metallicities (0.08 $\ge$ [\={Z}] $\ge$ $-$0.87) represented by the total sample. It indicates that metallicity does not have a significant effect on the FGLR distance, provided that many of the objects used in the analysis lie in the linear regime at higher flux-weighted gravities (log $g_F$ $\ge \sim$1.2). At lower flux-weighted gravities, it is unclear whether the curvature of the FGLR predicted by stellar evolution theory and its metallicity dependence is supported by the observational data. Further studies in this regime are needed to untangle this effect. 

\subsection{The Metallicity of NGC~3109}

\subsubsection{The Mass-Metallicity Relation}
\begin{figure}
\begin{center}
\includegraphics[scale=0.35]{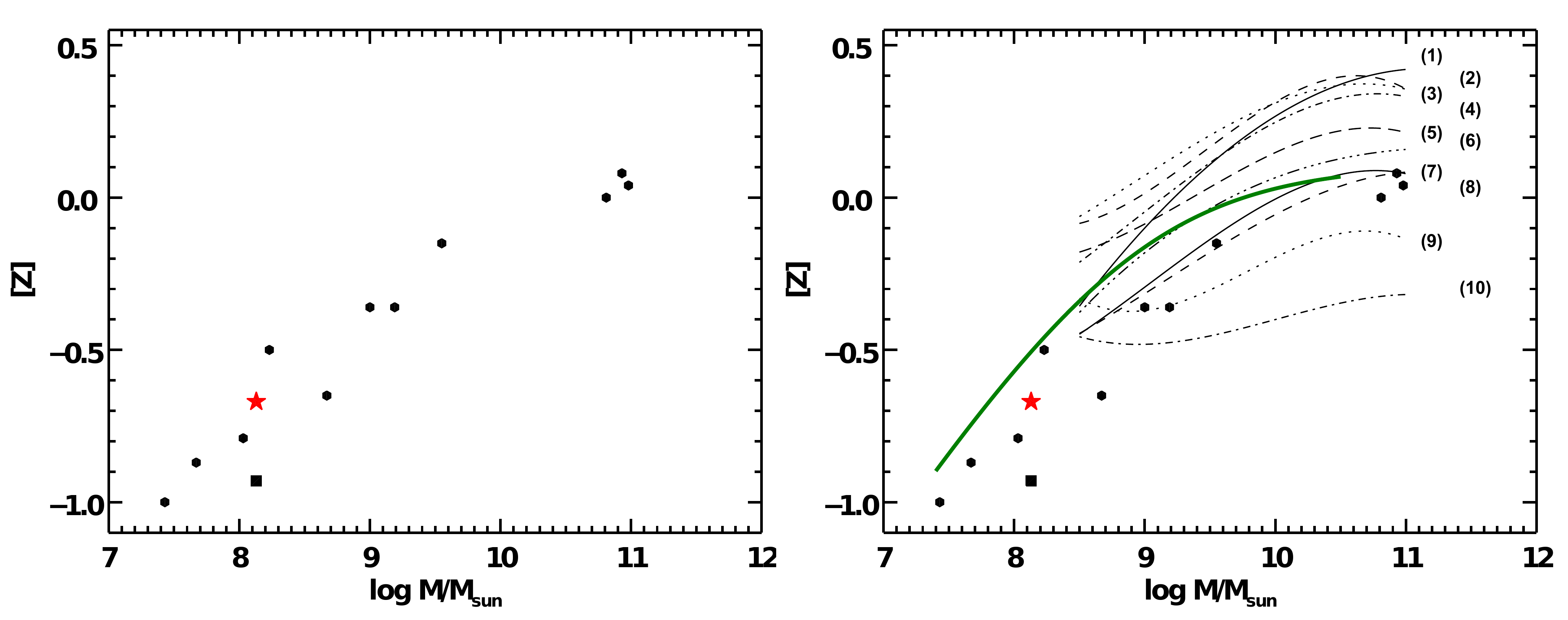}
\caption{\emph{Left:} The galaxy mass-metallicity relation as determined through studies of BSGs. The errors on these metallicities are typically 0.1-0.15 dex. Our derived metallicity for NGC~3109 is denoted by the red star, while the black square represents the metallicity of NGC~3109 determined by E07. \emph{Right:} A comparison of the BSG-based galaxy mass-metallicity relation with those determined from SDSS galaxies. The black lines show the 10 relations given by \citet{Ke08}: (1) solid, \citet{T04}; (2) dashed, \citet{Z94}; (3) dotted, \citet{KK04}; (4) dash-dotted, \citet{Ke02}; (5) long-dashed, \citet{Mc91}; (6) dash-triple-dotted, \citet{De02}, (7) solid, (\citeauthor{PP04}  2004; using [O III]/H$\beta$ and [N II] / H$\alpha$); (8) dashed, (\citeauthor{PP04} 2004; using [N II] / H$\alpha$); (9) dotted, \citet{Pil01}; and (10) dotted, \citet{Pil05}. The thick green line represents the recent relation by \cite{A13}, made using direct metallicity measurements of stacked SDSS spectra.}
\label{MM}
\end{center}
\end{figure}

Because of the systematic uncertainties affecting galaxy metallicities determined using strong-line measurements of HII regions, Kudritzki et al. (2012) compiled a new galaxy mass-metallicity relation only using objects whose metallicities have been derived through quantitative spectroscopy of BSGs. While currently containing only 12 galaxies (Table \ref{BSGMM_relation}), efforts are underway to expand this sample in order to test existing strong line calibrations and their subsequent relations. In addition, the dwarf galaxies studied are an important probe of the low-mass regime of the mass-metallicity relation, which is important for our understanding of galaxy formation and evolution \citep{Lee06}. NGC~3109 is one of these dwarf galaxies, and we adjust its position based on the average metallicities found in our study (Fig. \ref{MM}). Our result appears to be in better agreement than E07's result with the others in the sample, although the BSG-derived relation is still too sparse to make a definitive assessment at this point. 

Following \citet{K12}, we compare the BSG-based galaxy mass-metallicity relation to those derived using HII regions of star-forming galaxies in SDSS (Fig. \ref{MM}). We include the 10 mass-metallicity relations from \citet{Ke08}, each one employing a different strong-line calibration on the the same set of $\sim$25,000 galaxy spectra. The result is 10 distinctly different relations (demonstrating the uncertainties surrounding the calibrations), several of which are clearly discrepant with the BSG results. We also include the recent SDSS mass-metallicity relation presented by \citet{A13}, which has two major advantages: (1) galaxy metallicities are measured directly using auroral lines, eliminating the need for strong-line calibrations, and (2) the relation extends down to galaxy masses of log (M/M$_{\odot}$) = 7.4, equivalent to the lowest mass galaxies studied using BSGs and significantly beyond the lower limits of the \citet{Ke08} relations. The authors achieve this by stacking spectra of galaxies with similar masses in order to obtain the S/N required to measure the faint auroral lines, creating an average spectrum for each mass bin from which metallicity can be directly measured. The \citet{A13} relation appears to be qualitatively similar to that of the BSGs, though their metallicities appear to be systematically higher than ours. There could be several explanations for this, such as small-sample statistics in the BSG sample and differences in how the galaxy masses are determined. Clearly this requires further investigation. 

\subsubsection{A Metallicity Discrepancy?}

\begin{figure}
\begin{center}
\includegraphics[scale=0.55]{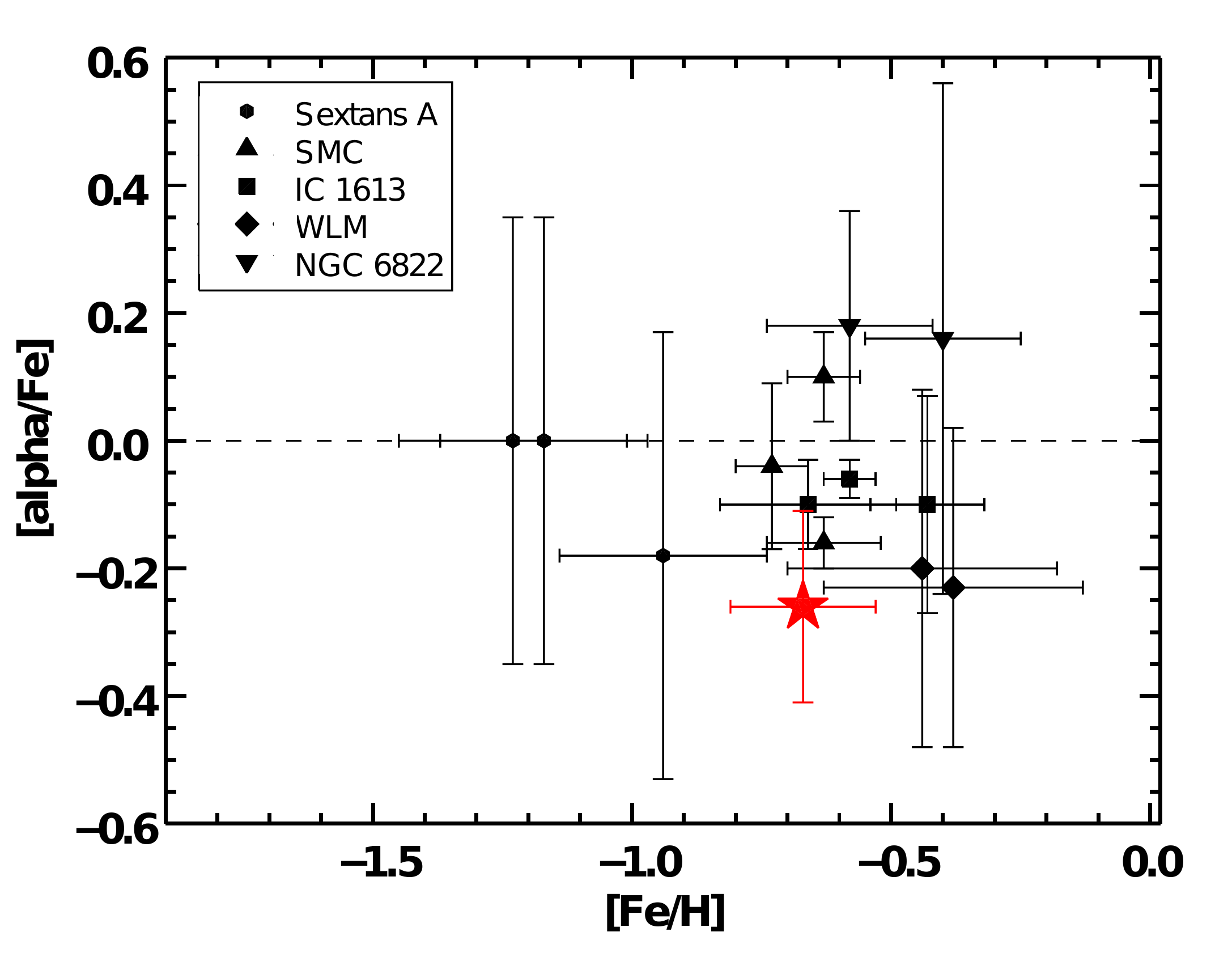}
\caption{[$\alpha$/Fe] vs. [Fe/H] derived from supergiants for different dwarf Irregular galaxies, adapted from \citet{T09}. Along with NGC~3109 (this study, denoted by the red star), the metallicities of Sextans A \citep{K04}, SMC \citep{V99, H97, L98}, IC 1613 \citep{Taut07}, WLM \citep{V03}, and NGC 6822 \citep{V01} are presented. The dotted line represents a $\alpha$/Fe ratio equal to solar. It is worthwhile to note that individual stellar Mg and Fe abundances are used to trace [$\alpha$/Fe] in these studies, while for NGC~3109 we compare the average oxygen abundance of \citet{E07} to our Fe-group metallicity.}
\label{alphairon}
\end{center}
\end{figure}

The average metallicity of our late-B and early-A sample ([\={Z}] = $-$0.67 $\pm$ 0.13) is notably higher than the metallicity found from early-B BSGs by E07 ([\={Z}] = $-$0.93 $\pm$ 0.07). This seems to indicate a systematic difference between the two samples. The metallicities in our study are derived from different elements than in E07;  we derive our metallicities primarily from Fe-group elements, such as Cr and Fe, and assuming a solar abundance pattern, while E07 derive their metallicities mostly from a set of isolated O lines (also one Mg and a few Si lines). The discrepancy between these results would seem to suggest that the Fe-group elements are enhanced compared to the $\alpha$-group elements in NGC~3109. This is unusual because, at least for older stellar populations, the $\alpha$-group elements are typically enhanced at lower metallicities \citep[see review by][]{Mc97}.  However, low $\alpha$/Fe ratios for galaxies as metal-poor as NGC~3109 is not unheard of. Studies of low-metallicity dwarf irregular galaxies have revealed a wide range of $\alpha$/Fe ratios, some lower than what is observed in the Milky Way (Fig. \ref{alphairon}, see review by \citeauthor{T09} 2009). This is thought to reflect the different star formation histories of these galaxies, though the exact mechanisms are not yet known. We note that the assumption of a solar $\alpha$/Fe ratio used in our model grid does not severely affect our metallicity determination, which is dominated by the Fe-group metal lines. 

There are several ways to investigate this interesting possibility further. A re-analysis of our sample using a significantly extended grid of model atmospheres with varying $\alpha$/Fe ratios and focusing on spectral windows with $\alpha$-element lines would require a substantial computational effort and careful testing of the accuracy of the method. A more direct and conventional method of determining the $\alpha$/Fe ratio would be to conduct a detailed spectroscopic analysis of high quality/resolution spectra of a few of our targets. At the distance of NGC~3109 and the brightness of our targets such an approach would be feasible, but would require a substantial amount of 8m-telescope time. 

\section{Conclusions}

We analyze low-resolution ($\sim$4.5~\AA) spectra of 12 late-B and early-A BSGs in NGC~3109, obtaining the effective temperatures, gravities, metallicities, reddening, and luminosities of these objects. We use a modified method of analysis that does not use the Balmer jump to break the temperature-metallicity degeneracy. Instead, we employ a $\chi^2$-based approach that takes advantage of the fact that spectral lines from different atomic species and from different excited levels react differently to changes of temperature and metallicity and use this to constrain stellar parameters. A test analysis of SMC spectra is found to produce parameters consistent with high-resolution analyses, attesting to the accuracy of our technique. A disadvantage of this method is that we must make assumptions regarding the He abundances of our objects, which we cannot determine from our analysis alone. Thus we consider two sets of model atmospheres, one assuming a ``normal'' He abundance (based on averages from BSG studies in the MW) and another assuming an enhanced He abundance, and take the average metallicity and FGLR distance as our final results for NGC 3109. Fortunately, the adopted He abundance only has a small impact on the derived metallicities, with the temperatures, gravities, reddening, and luminosities being very similar for both sets of models. 

From our sample, we find the average Fe-group metallicity of NGC~3109 to be [\={Z}] = $-$0.63 $\pm$ 0.13 in the normal He abundance case and [\={Z}] = $-$0.71 $\pm$ 0.14 in the enhanced He abundance case, resulting in our adopted metallicity of  [\={Z}] = $-$0.67 $\pm$ 0.13. Even in the enhanced He abundance case, this result is higher than the oxygen-based metallicity obtained by \citet{E07}, who find an average value of [\={Z}] = $-$0.93 $\pm$ 0.07 based on an analysis of 8 early-B BSGs. This may indicate a sub-solar $\alpha$/Fe ratio in the galaxy. We adjust the position of NGC~3109 on the BSG-based galaxy mass-metallicity relation presented in \citet{K12},  and find that the relation compares well (in a qualitative sense) to the recent mass-metallicity relation of \citet{A13} based on auroral line metallicity measurements of star-forming galaxies in SDSS. Interestingly, the metallicities of the \citet{A13} relation appear to be systematically higher than those in the BSG relation, an inconsistency which requires further investigation. 

We combine our results with the BSGs analyzed by \citet{E07} to determine the Flux-weighted Gravity-Luminosity Relation (FGLR) of NGC~3109. We find the FGLR to be almost identical to those found in other galaxies, demonstrating the consistency of the relation across a wide range of galaxy masses and metallicities. We obtain an FGLR distance modulus of $\mu$ = 25.55 $\pm$ 0.09 (1.27 Mpc) that is effectively independent of the adopted stellar He abundances. This result is in good agreement with distances found in Cepheid variable and TRGB studies, serving as an independent confirmation of these values. The consistency between our FGLR distance and the Cepheid distances of \citet{Pi06} and \citet{So06} suggests that the Cepheid period-luminosity (PL) relation is not strongly affected by metallicity, since these studies adopt PL slopes derived for higher metallicities ([Z]~=~$-$0.3). This study is an additional example of the great value of blue supergiants as independent distance and metallicity indicators in nearby galaxies.

\acknowledgments
The authors would like to thank J.P. Henry for useful discussions regarding the statistics used in this study and the anonymous referee for the extremely careful review of the manuscript that greatly improved the paper. This work was supported by the National Science Foundation under grant AST-1008798 to R.P.K. and F.B. R.P.K. acknowledges the support and hospitality of the University Observatory Munich and the MPA Garching, where part of this work was carried out. W.G. and G.P. are grateful for support from the BASAL Centro de Astrofisica y Tecnologias Afines (CATA) PFB-06/2007. Support from the Ideas Plus grant of the Polish Ministry of Science and Higher Education is also acknowledged.

Facilities: \facility{VLT:Antu (FORS2)}

\bibliography{ApJ_3109}


\clearpage


\begin{deluxetable}{c c c c c c c c}
\tablewidth{0pt}
\tabletypesize{\footnotesize}
\tablecaption{Late-B and early-A Type BSGs Analyzed in This Study}
\tablehead{
\colhead{ID\tablenotemark{a}} & \colhead{$\alpha$ (J2000.0)\tablenotemark{a}} & \colhead{$\delta$ (J2000.0)\tablenotemark{a}} & \colhead{Sp.T.\tablenotemark{a}} & \colhead{V (mag)\tablenotemark{b}} & \colhead{V $-$ I$_c$ (mag)\tablenotemark{b}} & V$_{rad}$ (km s$^{-1}$)\tablenotemark{a} & \colhead{Ave. Spectral S/N}}
\startdata
4 & 10 03 16.74 & $-$26 09 22.90 & B9 Ia & 18.36 & 0.14 & 454 &145\\
5  & 10 03 05.37 & $-$26 08 56.58 & B8 Ia & 18.54 & 0.05 & 411 & 123\\  
6 & 10 02 52.99 & $-$26 09 51.65 & B8 Ia & 18.62 & 0.05 & 370 & 157\\
13 & 10 03 08.51 & $-$26 09 57.31 & A0 Iab & 19.02 & 0.16 & 405 & 132\\
17 & 10 03 20.21 & $-$26 06 44.62 & A3 II & 19.17 & 0.26 & 447 & 82\\
21 & 10 03 09.96 & $-$26 08 27.11 & A0 Iab & 19.33 & 0.15 & 397 & 54\\
23 & 10 03 19.95 & $-$26 09 55.01 & A1 Iab & 19.44 & 0.14 & 453 & 87\\
25 & 10 03 22.95 & $-$26 10 30.91 & A0 Iab & 19.47 & 0.11 & 429 & 91\\
29 & 10 03 12.50 & $-$26 10 14.81 & B8 Ia & 19.54 & 0.02 & 421 & 90\\
30 & 10 02 55.53 & $-$26 09 54.82 & A0 Iab & 19.54 & $-$0.01 & 400 & 104\\
32 & 10 03 03.41 & $-$26 08 46.06 & A3 II & 19.57 & 0.25 & 370 & 80\\
40 & 10 03 14.72 & $-$26 09 57.40 & B8 Ib & 19.81 & $-$0.05 & 430 & 78
\enddata
\tablenotetext{a}{\citet{E07}}
\tablenotetext{b}{Pietrzy\'{n}ski et al. (2006)}
\label{Myobjects}
\end{deluxetable}

\begin{deluxetable}{c c c}
\tablewidth{0pt}
\tabletypesize{\footnotesize}
\tablecaption{Adopted He Abundances for Model Grid}
\tablehead{
\colhead{[Z]} &  \colhead{} & \colhead{$y$}}
\startdata
$-$1.30 & &0.08 \\
$-$1.15 & &0.08\\  
$-$1.00 & &0.08  \\
$-$0.85 & &0.09 \\
$-$0.70 & &0.09\\
$-$0.60 & &0.09  \\
$-$0.50 & &0.09  \\
$-$0.40 & &0.10  \\
$-$0.30 & &0.10 \\
$-$0.15 & &0.11  \\
 0.00 & &0.12 \\
 0.15 & &0.12 \\
 0.30 & &0.13 \\
 0.50 & &0.14   
\enddata
\tablenotetext{*}{$y = \frac{n_{He}}{n_{H} + n_{He}}$}
\label{Normal_He}
\end{deluxetable}

\begin{deluxetable}{c c c c c c}
\tablewidth{0pt}
\tabletypesize{\footnotesize}
\tablecaption{Spectral Windows}
\tablehead{
\colhead{Window ID} & \colhead{$\lambda$ Range (\AA)} & \colhead{Atomic Species} }
\startdata
1\tablenotemark{a} & 5220-5350 & Fe II  \\
2 & 5141-5191 & Fe II\\
3 & 4981-5081 & Fe II/He I (blend), Si II, S II  \\
4 & 4898-4847 & Fe II, He I  \\
5 & 4497-4643 & Fe II, Cr II  \\
6 & 4457-4497 & Mg II/He I  \\
7 & 4377-4424 & Fe II, He I \\
8\tablenotemark{a} & 4195-4332 & Fe II, Ti II  \\
9 & 4117-4196 & Si II,  Fe II\\  
10\tablenotemark{b} & 4018-4036 & He I \\
\enddata
\tablenotetext{a}{Only for early-A BSGs}
\tablenotetext{b}{Only for late-B BSGs}
\label{Spectral_win}
\end{deluxetable}

\begin{deluxetable}{c c c c c c | c c c}
\tablewidth{0pt}
\tabletypesize{\footnotesize}
\tablecaption{SMC Test Results}
\tablehead{
 & \multicolumn{5}{c}{Schiller (2010)\tablenotemark{a}} & \multicolumn{3}{c}{Our Analysis\tablenotemark{b}} \\
ID & Sp.T. & T$_{eff}$ (K) & log \emph{g} (cgs) & [Z] & $y$ & T$_{eff}$ (K) & log \emph{g} (cgs) & [Z] \\
}
\startdata
AV20 & B8 Ia & 8700 $\pm$ 50 & 1.10 $\pm$ 0.05 & $-$0.68 $\pm$ 0.09 & 0.13 & 9100$^{250}_{450}$ & 1.23$^{0.10}_{0.18}$ & $-$0.65$^{0.08}_{0.25}$ \\
 &  &  & & & & 9150$^{200}_{450}$ & 1.25$^{0.10}_{0.19}$ & $-$0.66$^{0.08}_{0.24}$ \\
 \hline
 AV76 & A0 Ia & 10250 $\pm$ 500 & 1.30 $\pm$ 0.10 & $-$0.65 $\pm$ 0.04 & 0.12 & 10500$^{200}_{280}$ & 1.34$^{0.08}_{0.09}$ & $-$0.66$^{0.20}_{0.08}$ \\
 &  &  &  &  & & 10500$^{125}_{160}$ & 1.34$^{0.08}_{0.08}$ & $-$0.70$^{0.13}_{0.21}$ \\
 \hline
 AV200 & B8 Ia & 12000 $\pm$ 500 & 1.70 $\pm$ 0.10 & $-$0.60 $\pm$ 0.18 & 0.12 & 12100$^{250}_{250}$ & 1.71$^{0.08}_{0.09}$ & $-$0.50$^{0.14}_{0.17}$ \\
 &  &  &  &  & & 11850$^{300}_{300}$ & 1.68$^{0.09}_{0.09}$ & $-$0.59$^{0.13}_{0.15}$ \\
\enddata
\tablenotetext{a}{Using high-resolution, high S/N spectra}
\tablenotetext{b}{Top results: Normal He models, Bottom Results: Enhanced He models}
\label{SMC_results}
\end{deluxetable}

\begin{deluxetable}{c c c c c c c c c c}
\tablewidth{0pt}
\tabletypesize{\footnotesize}
\tablecaption{Stellar Parameters of Late-B and Early-A Stars}
\tablehead{
ID & T$_{eff}$ & log \emph{g} & [Z] & $v_{t}$  & E(V-I$_c$) &E(B$-$V) & BC & m$_{bol}$ & R/R$_{25}$ \\
     &  (K) & (cgs) & (dex) & (km s$^{-1}$) & (mag) & (mag) & (mag) & (mag) & 
     }
\startdata
4 &  $11400^{250}_{250}$ & $1.71^{0.05}_{0.05}$ & $-0.59^{0.11}_{0.10}$ & 8 & 0.16 & 0.12 & $-$0.58 & 17.38 & 0.36  \\
 &  $11150^{250}_{250}$ & $1.67^{0.06}_{0.05}$ & $-0.70^{0.06}_{0.05}$ &  8 & 0.15 & 0.11 & $-$0.54 & 17.44 \\
 \hline
5 &  $12050^{250}_{250}$ & $1.73^{0.11}_{0.06}$ & $-0.48^{0.11}_{0.10}$ & 8 & 0.08 & 0.06 & $-$0.67 & 17.67 & 0.31 \\
 &  $11800^{250}_{250}$ & $1.69^{0.11}_{0.06}$ & $-0.55^{0.10}_{0.13}$ & 8 & 0.08 & 0.06 & $-$0.66 & 17.67   \\
 \hline
6 &  $12250^{250}_{250}$ & $1.78^{0.10}_{0.06}$ & $-0.67^{0.14}_{0.15}$ & 8 & 0.08 & 0.06 & $-$0.74 & 17.67 & 0.48 \\
 &  $11750^{270}_{250}$ & $1.72^{0.10}_{0.06}$ & $-0.85^{0.10}_{0.06}$ & 8 & 0.07 & 0.05 & $-$0.66 & 17.78   \\
 \hline
13 &  $9650^{100}_{100}$ & $1.80^{0.10}_{0.06}$ & $-0.66^{0.09}_{0.06}$ & 5 & 0.15 & 0.11 & $-$0.25 & 18.40 &  0.22 \\
 &  $9600^{130}_{130}$ & $1.78^{0.11}_{0.06}$ & $-0.68^{0.10}_{0.03}$ & 4 & 0.15 & 0.11 & $-$0.24 & 18.41   \\
\hline
17 &  $8150^{130}_{130}$ & $1.45^{0.15}_{0.16}$ & $-0.69^{0.10}_{0.18}$ & 4 & 0.20 & 0.15 & 0.03 & 18.70 & 1.64  \\
 &  $8150^{130}_{130}$ & $1.45^{0.15}_{0.16}$ & $-0.69^{0.09}_{0.18}$ & 4 & 0.20 & 0.15 & 0.03 & 18.69    \\
\hline
21 &  $9000^{470}_{400}$ & $1.72^{0.23}_{0.27}$ & $-0.57^{0.26}_{0.21}$ & 4 & 0.12 & 0.09 & $-$0.12 & 18.91 & 0.61 \\
 &  $9000^{380}_{450}$ & $1.72^{0.21}_{0.30}$ & $-0.57^{0.26}_{0.21}$ & 4 & 0.12 & 0.10 & $-$0.12 & 18.89    \\
\hline
23 &  $8750^{250}_{350}$ & $1.75^{0.19}_{0.27}$ & $-0.61^{0.23}_{0.34}$ & 4 & 0.11 & 0.08 & $-$0.08 & 19.09 & 0.45  \\
 &  $8750^{150}_{150}$ & $1.75^{0.21}_{0.30}$ & $-0.62^{0.24}_{0.34}$ & 4 & 0.11 & 0.09 & $-$0.08 & 19.08    \\
\hline
25 &  $8750^{130}_{130}$ & $1.69^{0.11}_{0.11}$ & $-0.99^{0.13}_{0.18}$ & 4 & 0.07 & 0.05 & $-$0.10 & 19.20 & 0.68 \\
 &  $8750^{150}_{150}$ & $1.69^{0.13}_{0.13}$ & $-1.00^{0.13}_{0.18}$ & 4 & 0.08 & 0.06 & $-$0.10 & 19.18    \\
\hline
29 &  $12800^{380}_{480}$ & $2.17^{0.09}_{0.10}$ & $-0.69^{0.20}_{0.22}$ & 7 & 0.09 & 0.07 & $-$0.83 & 18.48 & 0.40 \\
 &  $12250^{320}_{300}$ & $2.09^{0.09}_{0.09}$ & $-0.84^{0.25}_{0.19}$ & 6 & 0.07 & 0.05 & $-$0.74 & 18.62    \\
\hline
30 &  $10500^{130}_{130}$ & $1.94^{0.08}_{0.06}$ & $-0.50^{0.13}_{0.15}$ & 4 & 0.00 & 0.00 & $-$0.40 & 19.13 & 0.42 \\
 &  $10500^{150}_{150}$ & $1.94^{0.08}_{0.06}$ & $-0.50^{0.13}_{0.22}$ & 4 & 0.01 & 0.01 & $-$0.40 & 19.12   \\
\hline
32 &  $8300^{50}_{210}$ & $1.70^{0.11}_{0.21}$ & $-0.64^{0.17}_{0.17}$ & 4 & 0.21 & 0.16 & 0.00 & 19.06 & 0.40 \\
 &  $8300^{100}_{220}$ & $1.70^{0.13}_{0.21}$ & $-0.64^{0.17}_{0.17}$ & 4 & 0.21 & 0.16 & 0.01 & 19.05  \\
\hline
40 &  $12850^{380}_{450}$ & $2.34^{0.16}_{0.12}$ & $-0.65^{0.29}_{0.32}$ & 4 & 0.03 & 0.02 & $-$0.83 &  18.91 & 0.32 \\ 
 &  $12350^{350}_{350}$ & $2.26^{0.16}_{0.12}$ & $-0.79^{0.28}_{0.28}$ & 6 & 0.02 & 0.01 & $-$0.75 &  19.02  
\enddata
\label{Stellarparams}
\tablenotetext{*}{Top entry for each star is the analysis using the normal He abundance model, second entry is the analysis using the enhanced He abundance model}
\end{deluxetable}

\begin{deluxetable}{c c c | c c }
\tablewidth{0pt}
\tabletypesize{\footnotesize}
\tablecaption{$\chi^2_{min}$ of the Best-Fit Models}
\tablehead{
 & \multicolumn{2}{c}{Norm. He} & \multicolumn{2}{c}{Enhanced He} \\
ID & $\chi^2_{min}$ & $\chi^2_{\nu}$ & $\chi^2_{min}$ & $\chi^2_{\nu}$ 
}
\startdata
4 & 289 & 1.096 & 300 & 1.136\\
5 & 315 & 1.291 & 267 & 1.094\\  
6 & 341 & 1.322 & 334 & 1.296\\
13 & 638 & 1.289 & 639 & 1.292 \\
17 & 450 & 1.142 & 450 & 1.142\\
21 & 396 & 0.887 & 398 & 0.890\\
23 & 546 & 1.207 & 545 & 1.206\\
25 & 471 & 1.019 & 471 & 1.019\\
29 & 260 & 1.011 & 259 & 1.007\\
30 & 539 & 1.340 & 541 & 1.346\\
 32 & 533 & 1.268 & 532 & 1.266\\
 40 & 237 & 0.976 & 239 & 0.983
\enddata
\tablenotetext{*}{$\chi^2_{\nu} = \frac{\chi^2_{min}}{\nu}$}
\label{Chimin_comp}
\end{deluxetable}

\begin{deluxetable}{c l c c c c c c }
\tablewidth{0pt}
\tabletypesize{\footnotesize}
\tablecaption{Stellar Parameters of Early-B type Stars\tablenotemark{*}}
\tablehead{
\colhead{ID} & \colhead{Sp.T.} & \colhead{T$_{eff}$ (K)\tablenotemark{a}} & \colhead{log \emph{g} (cgs)\tablenotemark{b}} & \colhead{[Z]\tablenotemark{c,d}} & \colhead{E(B$-$V)} & \colhead{BC} & \colhead{m$_{bol}$} 
}
\startdata
3 & B1 Ia & 23500 & 2.50 & $-$1.0 & 0.09 & $-$2.37 & 15.41 \\
7 & B0-1 Ia & 27000 & 2.90 & $-$0.9 & 0.00 & $-$2.68 & 16.08    \\
9 & B0.5 Ia & 25000 & 2.75 & $-$0.9 & 0.01 & $-$2.50 & 16.33   \\
11 & B0 I & 27500 & 3.05 & $-$0.9 & 0.13 & $-$2.73 & 15.75   \\
22 & B1 Ia & 22000 & 2.60 & $-$0.9 & 0.00 & $-$2.20 & 17.20    \\
27 & B2.5 Ia & 19000 & 2.40 & $-$0.9 & 0.12 & $-$1.84 & 17.25    \\
28 & B2.5 Ia & 19000 & 2.55 &$-$0.9 & 0.16 & $-$1.84 & 17.13    \\
37 & B2 Iab & 20500 & 2.70 & $-$1.1 & 0.05 & $-$2.03 & 17.53 
\enddata
\tablenotetext{*}{From \citet{E07}}
\tablenotetext{a}{T$_{eff}$ uncertainty: $\pm$ 1000K}
\tablenotetext{b}{$log g$ uncertainty: $\pm$ 0.10 dex}
\tablenotetext{c}{[Z] uncertainty: $\pm$ 0.2 dex}
\tablenotetext{d}{Based on oxygen, with a solar oxygen abundance of 12 + log(O/H) = 8.66 \citep{Asp05}}
\label{Stellarparams_E07}
\end{deluxetable}

\begin{deluxetable}{c c c c c c }
\tablewidth{0pt}
\tabletypesize{\footnotesize}
\tablecaption{Absolute Magnitudes, Luminosities, Radii, and Masses for Total Sample}
\tablehead{
\colhead{ID} & \colhead{M$_{bol}$\tablenotemark{*}} & \colhead{log (L/L$_{\odot}$)} & \colhead{R (R$_{\odot}$)} & \colhead{M$_{evol}$ (M$_{\odot}$)} & \colhead{M$_{spec}$ (M$_{\odot}$)} 
}
\startdata
3 & -10.12 & 5.94 $\pm$ 0.11 & 56 $\pm$ 8 & $44$ $\pm$ 7 & 37 $\pm$ 15  \\
4 & -8.13 & 5.16 $\pm$ 0.11 & 97 $\pm$ 13 & $19$ $\pm$ 2 & 18 $\pm$ 5  \\
5 & -7.84 & 5.04 $\pm$ 0.11 & 79 $\pm$ 11 & $18$ $\pm$ 2& 11 $\pm$ 4  \\
6 & -7.73 & 5.05 $\pm$ 0.11 & 76 $\pm$ 10 & $18$ $\pm$ 2 & 12 $\pm$ 4  \\
7 & -9.43 & 5.67 $\pm$ 0.11 & 31 $\pm$ 5 & $32$ $\pm$ 4 & 29 $\pm$ 11 \\
9 & -9.19 & 5.57 $\pm$ 0.11 & 33 $\pm$ 5 & $29$ $\pm$ 4 & 22 $\pm$ 9 \\
11 & -9.76 & 5.80 $\pm$ 0.11 & 35 $\pm$ 5 & $38$ $\pm$ 5 & 51 $\pm$ 20  \\
13 & -7.12 & 4.75 $\pm$ 0.12 & 84 $\pm$ 12 & $14$ $\pm$ 1 & 17 $\pm$ 6 \\
17 & -6.81 & 4.63 $\pm$ 0.12 & 103 $\pm$ 15 & $13$ $\pm$ 1 & 15 $\pm$ 7 \\
21 & -6.61 & 4.55 $\pm$ 0.12 & 77 $\pm$ 14 & $12$ $\pm$ 1 & 11 $\pm$ 8 \\
22 & -8.31 & 5.22 $\pm$ 0.11 & 28 $\pm$ 5 & $20$ $\pm$ 2 & 14 $\pm$ 6 \\
23 & -6.43 & 4.48 $\pm$ 0.12 & 75 $\pm$ 12 & $11$ $\pm$ 1 & 12 $\pm$ 8 \\
25 & -6.32 & 4.43 $\pm$ 0.12 & 71 $\pm$ 10 & $11$ $\pm$ 1 & 9 $\pm$ 4 \\
27 & -8.27 & 5.20 $\pm$ 0.11 & 37 $\pm$ 6 & $20$ $\pm$ 2 & 13 $\pm$ 6  \\
28 & -8.38 & 5.25 $\pm$ 0.11 & 39 $\pm$ 7 & $21$ $\pm$ 2 & 20 $\pm$ 9 \\
29 & -7.04 & 4.72 $\pm$ 0.12 & 46 $\pm$ 7 & $13$ $\pm$ 1 & 12 $\pm$ 4 \\
30 & -6.38 & 4.46 $\pm$ 0.12 & 51 $\pm$ 7 & $11$ $\pm$ 1 & 8 $\pm$ 3 \\
32 & -6.46 & 4.49 $\pm$ 0.12 & 84 $\pm$ 13 & $11$ $\pm$ 1 & 14 $\pm$ 8  \\
37 & -7.99 & 5.09 $\pm$ 0.11 & 28 $\pm$ 5 & $18$ $\pm$ 2 & 14 $\pm$ 6 \\
40 & -6.61 & 4.55 $\pm$ 0.12 & 38 $\pm$ 6 & $12$ $\pm$ 1 & 12 $\pm$ 6  
\enddata
\tablenotetext{*}{Calculated assuming our FGLR distance modulus of $\mu$ = 25.55}
\label{Masscomp_table}
\end{deluxetable}

\begin{deluxetable}{c c}
\tablewidth{0pt}
\tabletypesize{\footnotesize}
\tablecaption{Position Parameters of NGC~3109\tablenotemark{a}}
\tablehead{
\colhead{Parameter} &  \colhead{Value} 
}
\startdata
Central $\alpha$ & 10h 03m 06s \\
Central $\delta$ & $-$26$^{\circ}$ 09' 32"\\  
$i$ & 75 $\pm$ 2$^{\circ}$  \\
Major Axis PA & 93 $\pm$ 2$^{\circ}$ \\
R$_{25}$ & 14.4' 
\enddata
\tablenotetext{a}{\citet{J90}}
\label{3109_pos}
\end{deluxetable}

\begin{deluxetable}{c c c c c}
\tablewidth{0pt}
\tabletypesize{\footnotesize}
\tablecaption{BSG Mass-Metallicity Relationship}
\tablehead{
\colhead{Galaxy} & \colhead{log (M/M$_{\odot}$)} & \colhead{[Z]} & \colhead{Source (mass)} & \colhead{Source [Z]} } 
\startdata
M81 & 10.93 & 0.08 & \tablenotemark{a} & \tablenotemark{b}  \\
M31 & 10.98 & 0.04 & \tablenotemark{c} & \tablenotemark{d,e}   \\
MW & 10.81 & 0.00 & \tablenotemark{f} & \tablenotemark{g}  \\
M33 & 9.55 & $-$0.15 & \tablenotemark{i}& \tablenotemark{h}  \\
NGC 300 & 9.00 & $-$0.36 & \tablenotemark{j} & \tablenotemark{k}    \\
LMC & 9.19 &  $-$0.36  &  \tablenotemark{i} &  \tablenotemark{l}     \\
SMC & 8.67 &  $-$0.65 &  \tablenotemark{i} &  \tablenotemark{m,n}     \\
NGC 6822 & 8.23 & $-$0.50 &  \tablenotemark{i} &  \tablenotemark{o} \\
NGC~3109 & 8.13 & $-$0.67 &  \tablenotemark{i} &  This study \\
IC 1613 & 8.03 & $-$0.79 & \tablenotemark{i} & \tablenotemark{p} \\
WLM & 7.67 & $-$0.87 & \tablenotemark{i} &  \tablenotemark{q} \\
Sex A & 7.43 & $-$1.00 &  \tablenotemark{i} &  \tablenotemark{r} 
\enddata
\tablenotetext{a}{\citet{dB08}}
\tablenotetext{b}{\citet{K12}}
\tablenotetext{c}{\citet{C09}}
\tablenotetext{d}{\citet{Tr02}}
\tablenotetext{e}{\citet{Sm01}}
\tablenotetext{f}{\citet{Sof09}}
\tablenotetext{g}{\citet{Pr08}}
\tablenotetext{h}{\citet{U09}}
\tablenotetext{i}{\citet{Woo08}}
\tablenotetext{j}{\citet{K87}}
\tablenotetext{k}{\citet{K08}}
\tablenotetext{l}{\citet{H07}}
\tablenotetext{m}{\citet{Sch10}}
\tablenotetext{n}{\citet{TL05}}
\tablenotetext{o}{\citet{V01}}
\tablenotetext{p}{\citet{B07}}
\tablenotetext{q}{\citet{U08}}
\tablenotetext{r}{\citet{K04}}
\label{BSGMM_relation}
\end{deluxetable}

\clearpage

\appendix
\section{Appendix: Extra Figures}

\begin{figure}
\begin{center}
\includegraphics[scale=0.40]{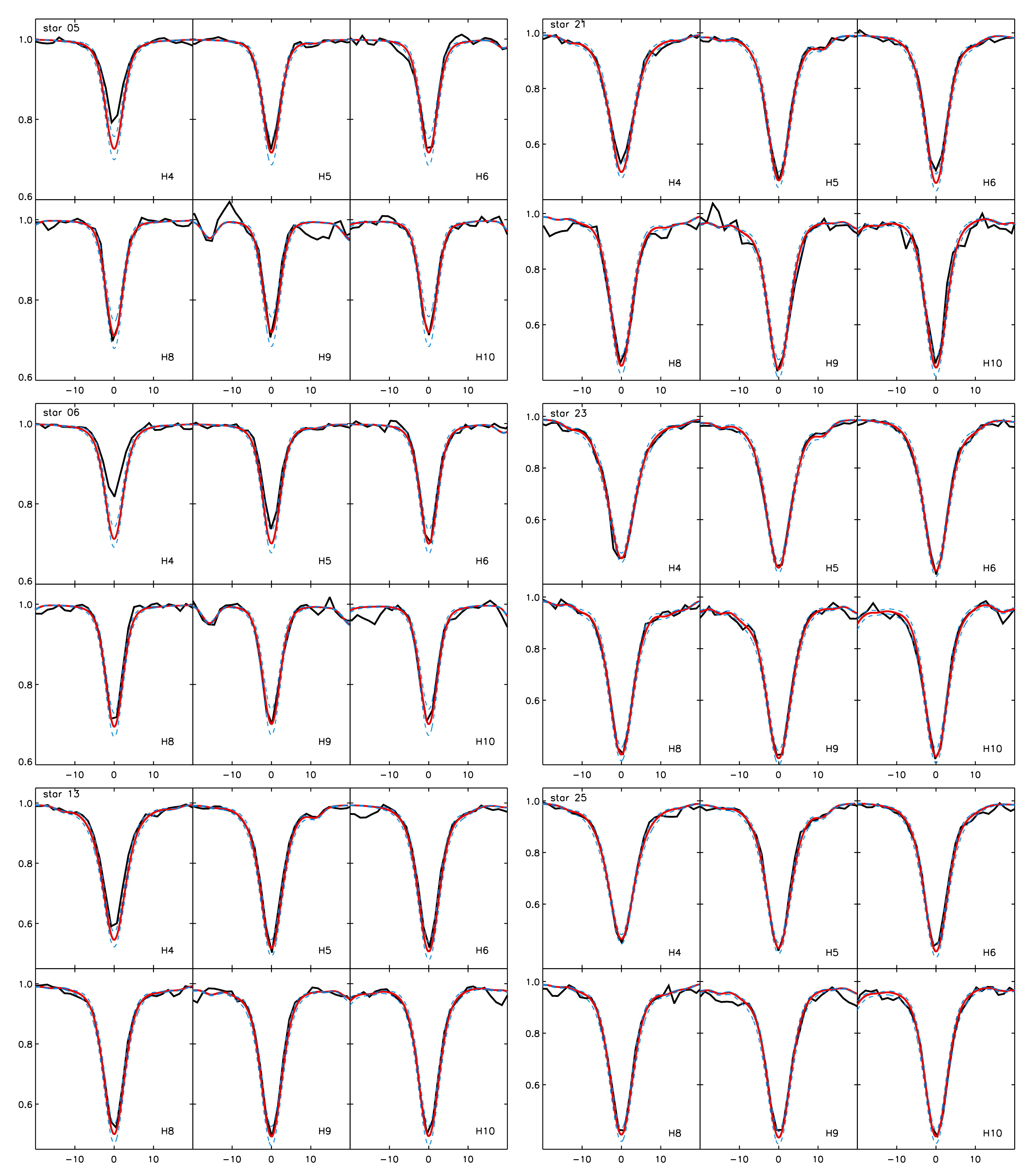}
\caption{Balmer line fits for the additional late-B and early-A type BSGs in this study. Observed spectrum is black line, best-fit model is red line (normal He abundance), and log g $\pm$ 0.1 dex is shown by the dashed blue lines. The left column contains stars 5, 6, and 13, while the right column contains stars 21, 23, and 25.}
\label{Append1}
\end{center}
\end{figure}

\begin{figure}
\begin{center}
\includegraphics[scale=0.40]{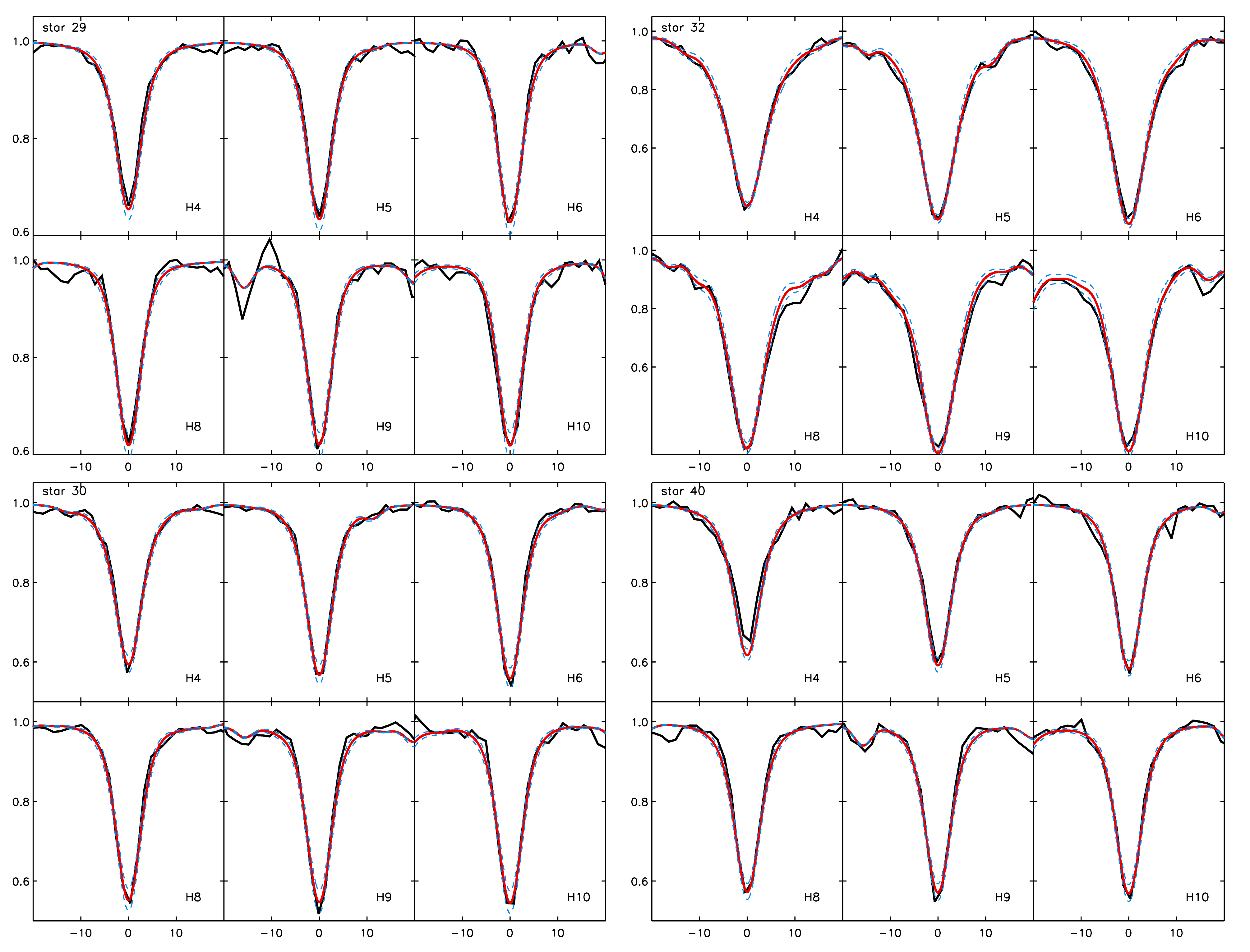}
\caption{Balmer line fits for the additional late-B and early-A type BSGs in this study. Observed spectrum is black line, best-fit model is red line (normal He abundance), and log g $\pm$ 0.1 dex is shown by the dashed blue lines. The left column shows stars 29 and 30 while the right column shows stars 32 and 40.}
\label{Append2}
\end{center}
\end{figure}

\begin{figure}
\begin{center}
\includegraphics[scale=0.40]{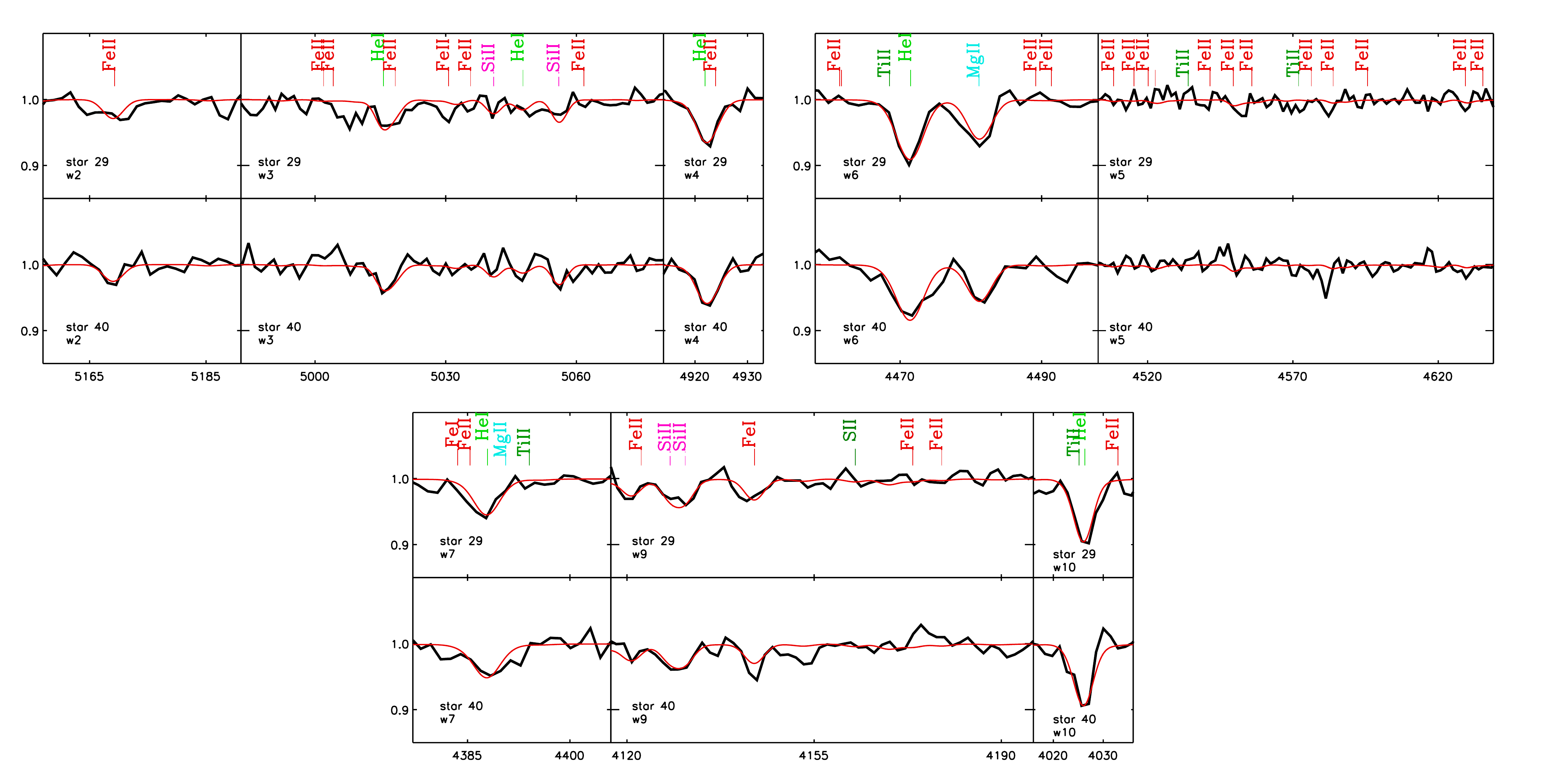}
\caption{Metal line fits for stars 29 and 40, two late-B BSGs in this study. Observed spectrum is in black, best-fit model (normal He abundance) in red. The enhanced He abundance fits are of similar quality.}
\label{Append3}
\end{center}
\end{figure}

\begin{figure}
\begin{center}
\includegraphics[scale=0.40]{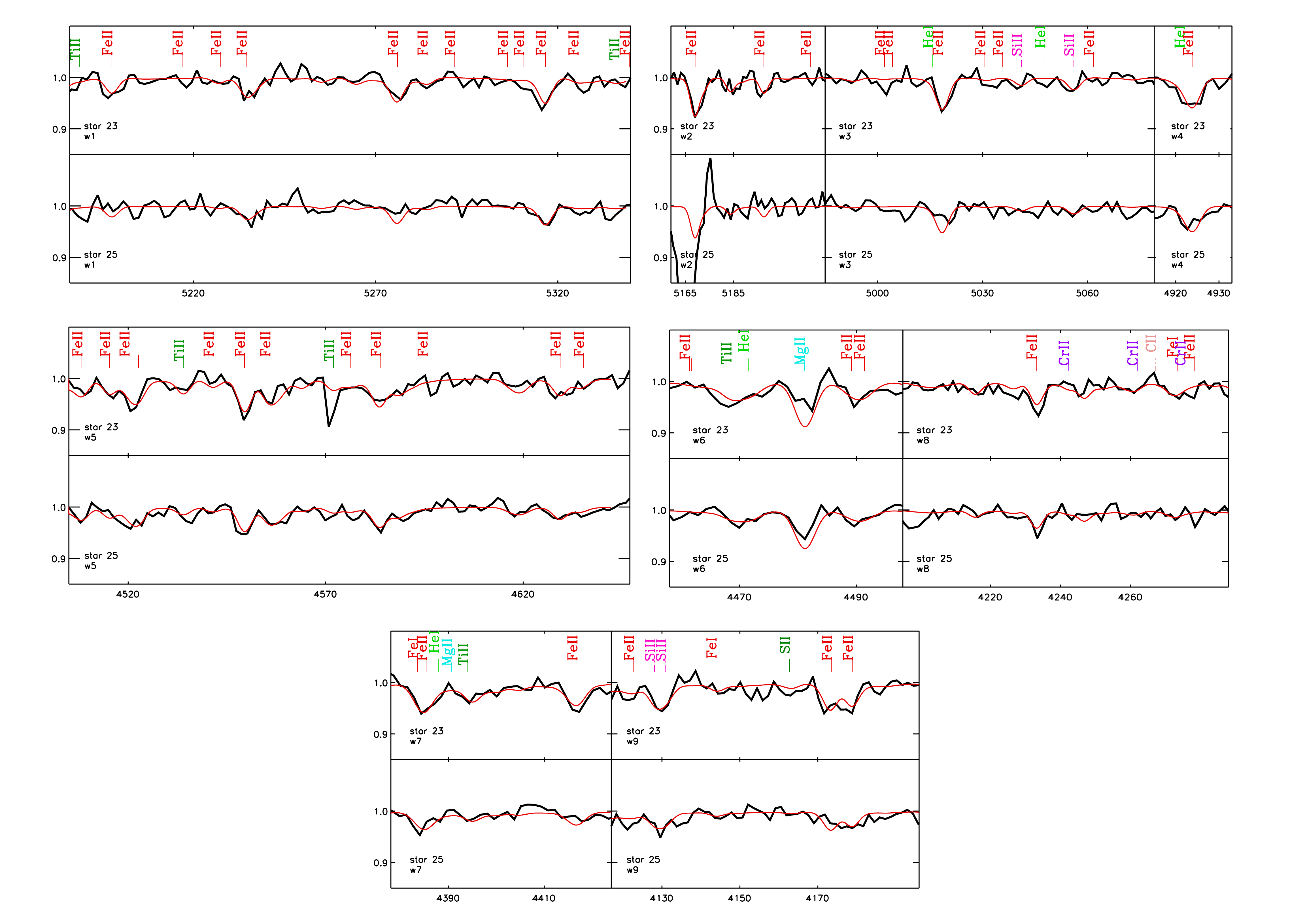}
\caption{Metal line fits for stars 23 and 25, two early-A BSGs in this study. Observed spectrum is in black, best-fit model (normal He abundance) in red. The enhanced He abundance fits are of similar quality. Note that star 25 has significantly weaker metal lines, indicating a lower metallicity as found by our analysis. }
\label{Append4}
\end{center}
\end{figure}


\begin{figure}
\begin{center}
\includegraphics[scale=0.35]{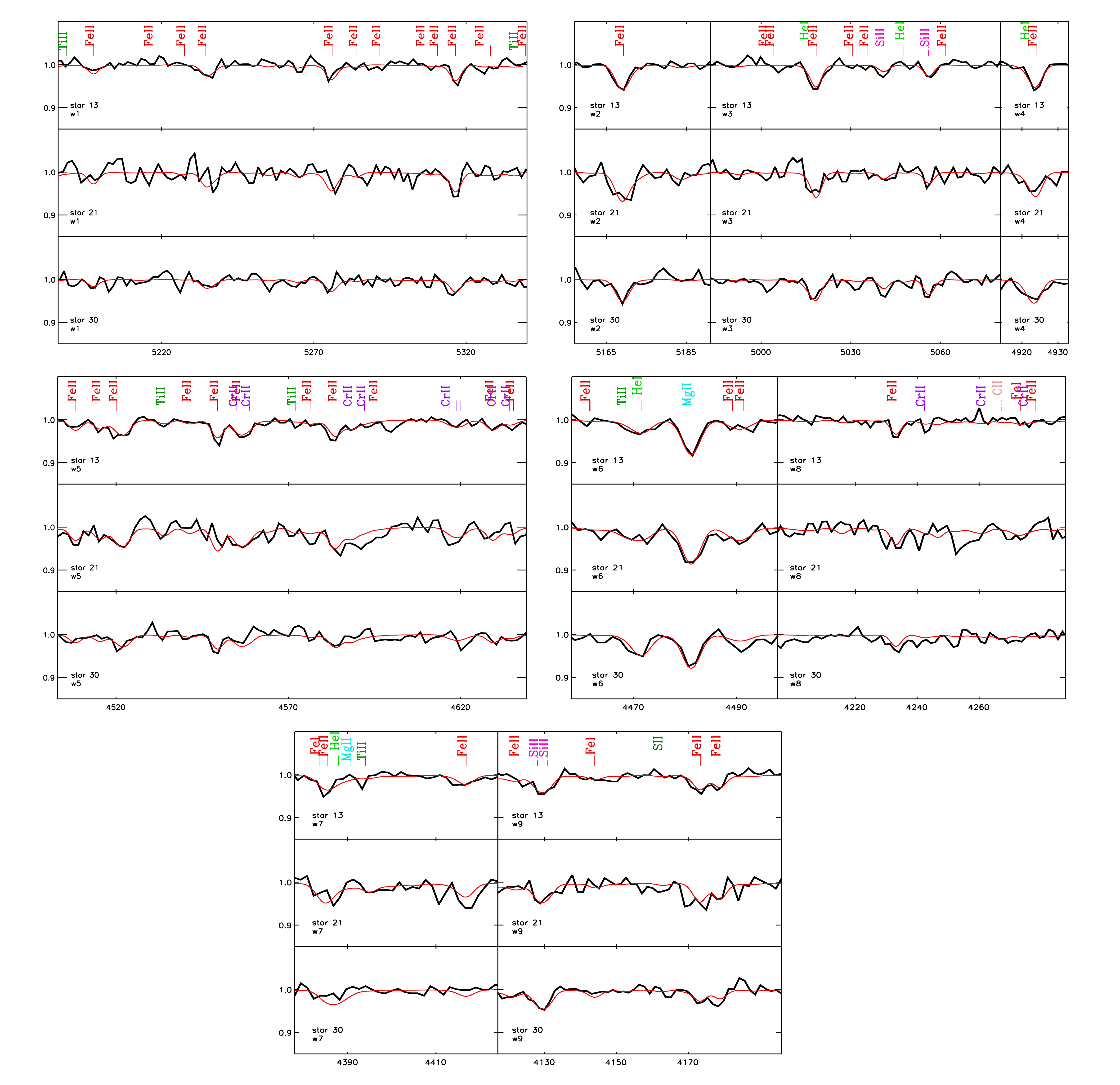}
\caption{Metal line fits for stars 13, 21, and 30, three early-A BSGs in this study. Observed spectrum is in black, best-fit model (normal He abundance) in red. The enhanced He abundance fits are of similar quality.}
\label{Append5}
\end{center}
\end{figure}

\begin{figure}
\begin{center}
\includegraphics[scale=0.25]{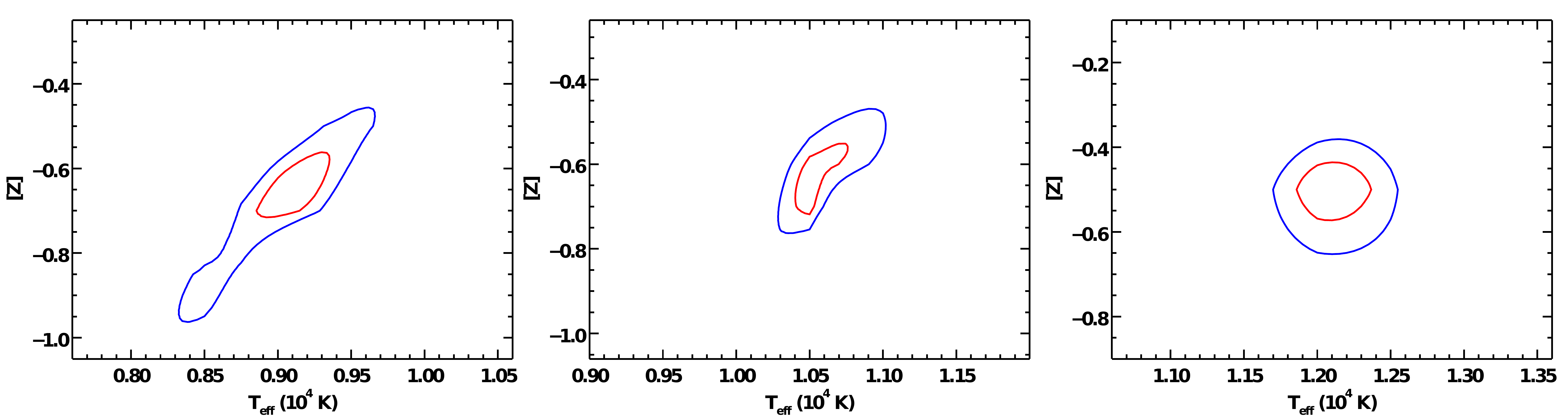}
\caption{Chi-square fit isocontours of the SMC test stars: AV20, AV76, and AV200 from left to right. The red line represents the 1$\sigma$ is contour while the blue line represents the 2$\sigma$ isocontour.}
\label{Append6}
\end{center}
\end{figure}

\begin{figure}
\begin{center}
\includegraphics[scale=1.0]{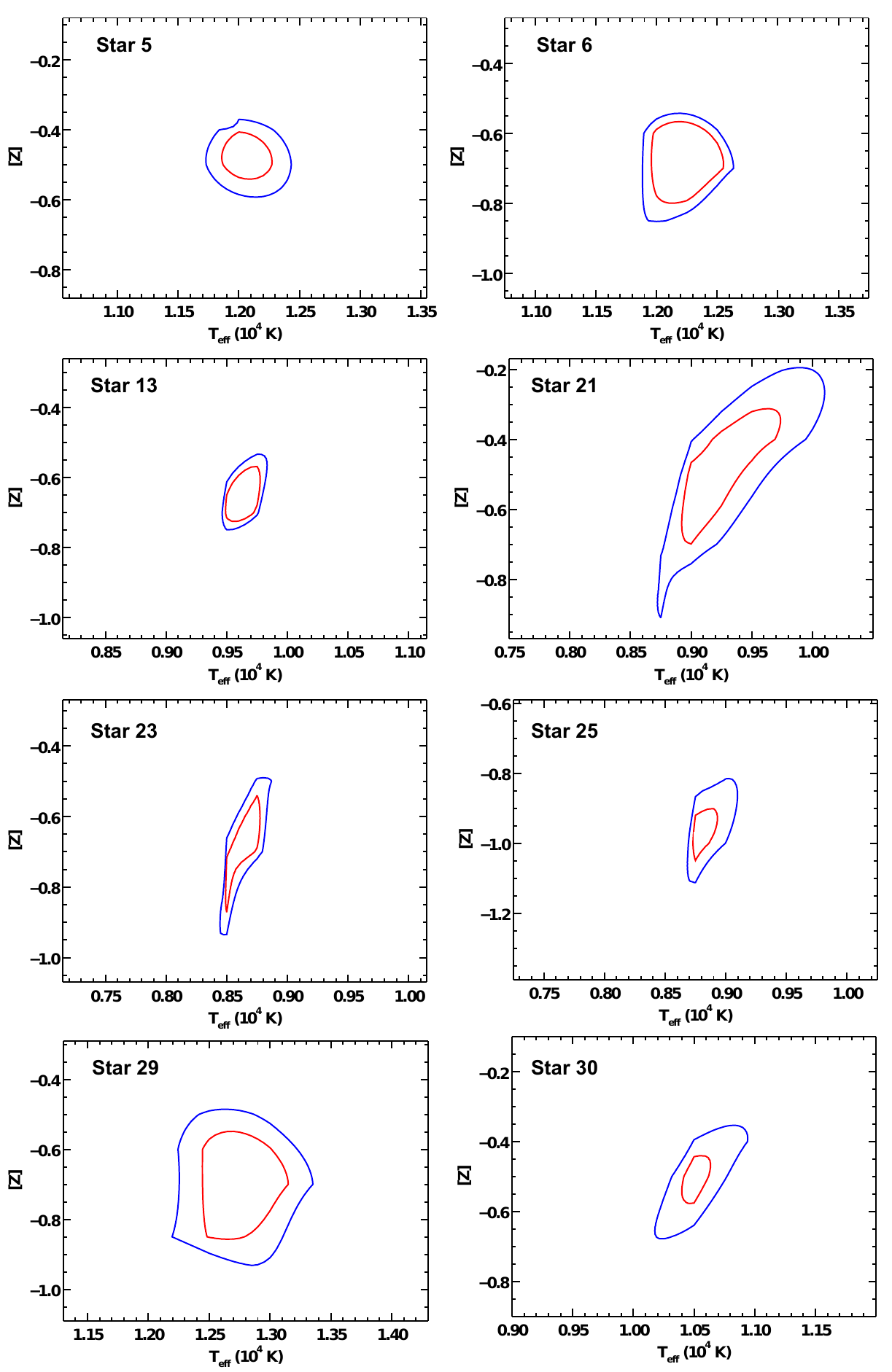}
\caption{Chi-square fit isocontours for the remaining stars in our study (stars 4 and 17 are shown in Fig. \ref{Analysis}). The red line represents the 1$\sigma$ is contour while the blue line represents the 2$\sigma$ isocontour.}
\label{Append6}
\end{center}
\end{figure}

\end{document}